\documentclass[journal]{IEEEtran}

\IEEEoverridecommandlockouts                              % This command is only
\usepackage{amssymb}
\usepackage{amsmath}
\usepackage{bm}
\usepackage{mathrsfs}
\usepackage{comment} 
\usepackage{algorithmic,algorithm}
\usepackage{shuffle}
\usepackage{graphicx} % for pdf, bitmapped graphics files
\usepackage{caption}
\usepackage{lipsum}
\usepackage{color}
\usepackage{enumerate}
\usepackage{amsfonts}
\usepackage{subfigure} 
\usepackage{array}
\usepackage{enumerate}
\usepackage{setspace}
\usepackage{multirow}
\DeclareSymbolFont{symbolsC}{U}{pxsyc}{m}{n}
\DeclareMathSymbol{\coloneqq}{\mathrel}{symbolsC}{"42}

\begin{document}

\title{\LARGE \bf
Privacy-Preserving Co-synthesis  Against Sensor-Actuator Eavesdropping Intruder} %Partially Observed  via Projections of Holons
\author{Ruochen Tai, Liyong Lin, Yuting Zhu and Rong Su
%\thanks{
%This work was partially supported by the National Natural Science Foundation of China under Grant
%Nos. 61374068, 61472295, and 61672400, the Recruitment Program of Global Experts, and the Science
%and Technology Development Fund, MSAR, under Grant Nos. 078/2015/A3 and 106/20156/A3
%(Corresponding author: Z. Li).

%D.~Wang is with the School of Electro-Mechanical Engineering, Xidian University, Xi'an 710071, China (e-mail: wdeguang1991@163.com).
%
%L.~Lin is with the Systems Control Group, Department of Electrical
%and Computer Engineering, University of Toronto, Toronto, ON M5S 3G4
%Canada (e-mail: liyong.lin@utoronto.ca).
%
%Z. Li is with the Institute of Systems Engineering, Macau University of Science and Technology, Taipa,
%Macau and also with the Key Laboratory of Electronic Equipment Structure Design, Ministry of
%Education, School of Electro-Mechanical Engineering, Xidian University, Xi'an 710071, China (e-mail: zhwli@xidian.edu.cn).
%
%W.~M.~Wonham is with the Systems Control Group, Department of Electrical
%and Computer Engineering, University of Toronto, Toronto, ON M5S 3G4,
%Canada (e-mail: wonham@control.utoronto.ca).}
%}
\thanks{The research of the project was supported by Ministry of Education, Singapore, under grant AcRF TIER 1-2018-T1-001-245 (RG 91/18).

The authors are affliated with Nanyang Technological University, Singapore. (Email: ruochen001@e.ntu.edu.sg; liyong.lin@ntu.edu.sg; yuting002@e.ntu.edu.sg; rsu@ntu.edu.sg).
(\emph{Corresponding author: Liyong Lin})}
}
\maketitle

\begin{abstract}
In this work, we investigate the problem of privacy-preserving supervisory control against an external passive intruder via co-synthesis of  dynamic mask,  edit function, and  supervisor for opacity enforcement and requirement satisfaction. We attempt to achieve the following goals: 1) the system secret cannot be inferred by the intruder, i.e., opacity of secrets against the intruder, and the existence of the dynamic mask and the edit function should not be discovered by the intruder, i.e.,  covertness of dynamic mask and edit function against the intruder; 2) the closed-loop system behaviors should satisfy some safety and nonblockingness requirement. We assume the intruder can eavesdrop both the sensing information generated by the sensors and the control commands issued to the actuators, and we refer to such an intruder as a sensor-actuator eavesdropping intruder. Our approach is to model the co-synthesis problem as a distributed supervisor synthesis problem in the Ramadge-Wonham supervisory control framework, and we propose an incremental synthesis heuristic to incrementally synthesize a dynamic mask, an edit function and a supervisor, which consists of three steps: 1) we first synthesize an ensemble $ME$ of dynamic mask and edit function to ensure the opacity and the covertness against a  sensor eavesdropping but command non-eavesdropping intruder, and marker-reachability; 2) we then decompose $ME$ into a dynamic mask and an edit function by using a constraint-based approach, with the help of a Boolean satisfiability (SAT) solver; 3) finally, we synthesize a supervisor to leak as little information as possible from the issued control commands to the intruder such that opacity and covertness can be ensured against the sensor-actuator eavesdropping intruder, and at the same time safety and nonblockingness requirement can be ensured. The effectiveness of our approach is illustrated on an example about the enforcement of location privacy for an autonomous vehicle.
\end{abstract}


\section{Introduction}
\label{intro}
With the continuous development of information and communication technology (ICT), in particular, the recent 5G-based Internet of Things (IoT) technologies, we are enjoying unprecedented connectivity around the world, including the ability to remotely monitor and control various systems, for example, cyber physical systems (CPS). However, the threat of cyber attacks, which may potentially cause significant system damage and even cost human lives, imposes great challenges when deploying sensor and actuator devices and designing defending strategies. As an attribute that expresses system security in a general language based theoretical framework \cite{Berard2015QuantifyOpacity}, opacity has been widely investigated in the context of discrete-event systems (DES), including its verification and  enforcement.

For the opacity verification problem, \cite{Saboori2007SBO}, \cite{Wu2013ComparativeOpacity}, \cite{Lin2011OpacityLBO}, \cite{Saboori2008VerifyISBO} - \cite{Yang2021NetworkedOpacity} develop various algorithms to verify different kinds of opacity notions, including 1) current-state opacity; 2) initial-state opacity; 3) initial-and-final-state opacity; 4) $K$-step opacity; 5) infinite-step opacity, in different setups, including standard DES, stochastic DES, and networked DES. Readers could refer to \cite{Jacob2016OverviewOpacity,Lafortune2018OverviewOpacity} for more details.

To enforce the opacity, three kinds of techniques are usually adopted, including 1) supervisory control  \cite{Takai2009SCT}-\cite{Dubreil2008JDESOpacitySCT}; 2) edit function \cite{Falcone2015OpacityRuntime}, \cite{Wu2014AutoInsert}-\cite{Falcone2013CDCRuntime}; 3) dynamic mask \cite{Cassez2012FMSDMask,Yin2019TACMask,Zhang2015TASEMask}. 
All of the above-mentioned works either synthesize a supervisor to restrict the system behavior or synthesize an edit function or a mask to ensure the opacity when the system behavior is not restricted. %However, a more urgent
A practically relevant problem arising from the need of resilient control is to synthesize privacy-preserving supervisory control strategies for opacity enforcement and requirement satisfaction at the same time, and this problem has not been investigated in the above-mentioned works.
For this problem, on one hand, we need to ensure the system behaviors satisfies some user-given requirement, which might conflict with the opacity property, by adopting the supervisory control, and, on the other hand, we need to ensure the opacity against the external intruder by utilizing an edit function or a dynamic mask. The supervisor is required to satisfy the given requirement in the presence of the edit function or the dynamic mask. And, ideally, the supervisor should also help to ensure the opacity, if an edit function or a dynamic mask is not sufficient.

To the best of our knowledge, the only work that has studied this closed-loop privacy-preserving control problem is the recent work \cite{tai2021privacy}, where an edit function is used. In~\cite{tai2021privacy}, the intruder is assumed to observe some sensor events, and
an edit function and a supervisor are co-synthesized to enforce the opacity,  the covertness\footnote{Covertness means that the existence of the edit function should not be discovered by the intruder; the intruder could always compare the online observations with the prior knowledge to detect  information inconsistency, and once some inconsistency is detected, it could discover the edit function.} and the  requirement satisfaction. Nevertheless, there exist some limitations in the setup of \cite{tai2021privacy}: 1) the intruder can only observe sensor information but cannot eavesdrop control commands issued by the supervisor; 
%Such an restriction lightens the difficulty for the edit function to enforce the opacity and remain covert since if the control commands can also be observed by the intruder, then it could predict what events could be fired based on the observed control command and the prior knowledge of the plant structure, which leaves less space for the edit function to alter the sensor information than that in the setup where the intruder could only observe sensor events. 
2) only an edit function is deployed to confuse the intruder to ensure the opacity, which might not work in some scenarios where the capability of the edit function is not powerful enough. 
%In this case, the dynamic mask, as an effective technique to enforce opacity, could be taken use of to turn off sensors at some specific and critical moments to compensate for the insufficient capability of the edit function such that the secret information flow leakage could be prevented. 
In this work, by employing a dynamic mask together with an edit function and a supervisor against the more powerful sensor-actuator eavesdropping intruder for opacity enforcement and requirement satisfaction, we solve a new and more challenging privacy-preserving supervisory control problem, The challenges are as follows:  
1) The reinforcement of the intruder's observation capability not only increases the difficulty for the dynamic mask and the edit function to enforce the opacity and remain covert, but also reduces the feasible solution space for the supervisor since the eavesdropped control commands could help the intruder on state estimation and identify the events that should not have been fired based on the prior knowledge of plant $G$.
%{\color{red} thinking about commenting out the following long sentence as it is obvious and user can easily understand: since if the control commands can also be observed, then the intruder could predict what events could be fired based on the observed control command and the prior knowledge of the plant structure, which leaves less space for the dynamic mask and edit function to alter the sensor information than that in the setup where the intruder could only observe sensor information}. On the other hand, the feasible solution space of the supervisor is also restricted since the eavesdropped command information could assist the intruder in inferring the system secret, and inappropriate control commands issued by the supervisor could help the intruder refine its state estimate and make it hard for the dynamic mask and the edit function to alter the sensor information for opacity enforcement, as they are also required to remain covert.  
2) The dynamic mask and the edit function should cooperate with each other to make up for each other's insufficient capability on observing and altering sensing information to enforce the opacity. Thus, in the synthesis of the dynamic mask (respectively, the edit function), not only its capability should be considered but also the edit function's (respectively, the dynamic mask's) capability should be taken into account. 
In this work, we shall overcome these challenges and the contributions are summarized as follows:
\begin{enumerate}[1.]
\setlength{\itemsep}{3pt}
\setlength{\parsep}{0pt}
\setlength{\parskip}{0pt}
    \item We investigate the problem of  privacy-preserving supervisory control against a sensor-actuator eavesdropping intruder via co-synthesis of dynamic mask, edit function, and  supervisor for opacity enforcement and requirement satisfaction, which is a more realistic and more challenging problem in the context of resilient control. In this work, we adopt a general setup for this privacy-preserving control problem: 1) the intruder could not only observe the sensing information but also observe the command information; 2) the observation capabilities of the dynamic mask, the edit function, the supervisor, and the intruder could all be different; 3) the capability of the dynamic mask and the edit function in altering the sensing information  could be different. This general setup has never been considered in previous works on opacity enforcement. Moreover, we also ensure the covertness of the synthesized dynamic mask and edit function.
    \item By formulating the system components as finite state automata, the problem of co-synthesizing dynamic mask, edit function, and supervisor is addressed by modelling it as a distributed supervisor synthesis problem in the Ramadge-Wonham supervisory control framework.
    \item By exploiting the structure of this privacy-preserving supervisory control problem, the modeled distributed supervisor synthesis problem is addressed by an incremental synthesis heuristic. In the proposed heuristic, we firstly synthesize an ensemble of the dynamic mask and the edit function to ensure the marker-reachability, the opacity and the covertness against a  sensor eavesdropping, but command non-eavesdropping intruder; then, we decompose the ensemble into a dynamic mask and an edit function by employing a reduction to the Boolean Satisfiability (SAT) Problem; and finally, a supervisor is synthesized based on the dynamic mask and the edit function to ensure the nonblockingness, opacity and covertness against the sensor-actuator eavesdropping intruder.
   Our method attempts to synthesize a local supervisor to ensure the marker-reachability first, which has more chances to generate a feasible solution than the existing incremental synthesis approaches which always synthesize a nonblocking local supervisor at each step.
    %To solve the decomposition problem, we reduce the original problem to a Boolean Satisfiability Problem (SAT) problem, which can be effectively solved by the SAT solver.
\end{enumerate}

This paper is organized as follows. In Section \ref{sec:Preliminaries}, we present some basic preliminaries which are needed in this work. In Section \ref{sec:Component Models of DES under Mask-Edit Function-Supervisor}, we introduce the component models, including the dynamic mask, the edit function, and the supervisor. Section \ref{sec:Co-Synthesis of Dynamic Mask, Edit Function, and Supervisor for Opacity Enforcement} proposes a heuristic method to co-synthesize a dynamic mask, an edit function, and a supervisor. An example is given to illustrate the effectiveness of the proposed method in Section \ref{sec:example}. Finally, conclusions are drawn in Section \ref{sec:conclusions}. 

%%%%%%%%%%%%%%%%%%%%%%%%%%%%%%%%%%%%%%%%%%%%%%%%%%%%%%%%%%%%%%%%%%%%%%%%%%%%%%%%

\section{Preliminaries}
\label{sec:Preliminaries}
Given a finite alphabet $\Sigma$, let $\Sigma^{*}$ be the free monoid over $\Sigma$ with the empty string $\varepsilon$ being the unit element and the string concatenation being the monoid operation. For a string $s$, $|s|$ is defined as the length of $s$. Given two strings $s, t \in \Sigma^{*}$, we say $s$ is a prefix substring of $t$, written as $s \leq t$, if there exists $u \in \Sigma^{*}$ such that $su = t$, where $su$ denotes the concatenation of $s$ and $u$. A language $L \subseteq \Sigma^{*}$ is a set of strings. The prefix closure of $L$ is defined as $\overline{L} = \{u \in \Sigma^{*} \mid (\exists v \in L) \, u\leq v\}$. 
%If $L = \overline{L}$, then $L$ is \emph{prefix-closed}. The concatenation of two languages $L_{a}, L_{b} \subseteq \Sigma^{*}$ is defined as $L_{a}L_{b} = \{s_{a}s_{b} \in \Sigma^{*}|s_{a} \in L_{a} \wedge s_{b} \in L_{b}\}$.
The event set $\Sigma$ is partitioned into $\Sigma = \Sigma_{c} \dot{\cup} \Sigma_{uc} = \Sigma_{o} \dot{\cup} \Sigma_{uo}$, where $\Sigma_{c}$ (respectively, $\Sigma_{o}$) and $\Sigma_{uc}$ (respectively, $\Sigma_{uo}$) are defined as the sets of controllable (respectively, observable) and uncontrollable (respectively, unobservable) events, respectively.  As usual, $P_{o}: \Sigma^{*} \rightarrow \Sigma_{o}^{*}$ is the natural projection defined such that: 1) $P_{o}(\varepsilon) = \varepsilon$, 2) $(\forall \sigma \in \Sigma) \, P_{o}(\sigma) = \sigma$, if $\sigma \in \Sigma_{o}$, otherwise, $P_{o}(\sigma) = \varepsilon$, 3) $(\forall s \in \Sigma^*, \sigma \in \Sigma) \, P_{o}(s\sigma) = P_{o}(s)P_{o}(\sigma)$.

A finite state automaton $G$ over $\Sigma$ is given by a 5-tuple $(Q, \Sigma, \xi, q_{0}, Q_{m})$, where $Q$ is the state set, $\xi: Q \times \Sigma \rightarrow Q$ is the (partial) transition function, $q_{0} \in Q$ is the initial state, and $Q_{m}$ is the set of marker states. 
We write $\xi(q, \sigma)!$ to mean that $\xi(q, \sigma)$ is defined and also view $\xi \subseteq Q \times \Sigma \times Q$ as a relation. $En_{G}(q) = \{\sigma \in \Sigma|\xi(q, \sigma)!\}$.
$\xi$ is also extended to the (partial) transition function $\xi: Q \times \Sigma^{*} \rightarrow Q$ and the transition function $\xi: 2^{Q} \times \Sigma \rightarrow 2^{Q}$ \cite{wonham2015supervisory}, where the later is defined as: for any $Q' \subseteq Q$ and any $\sigma \in \Sigma$, $\xi(Q', \sigma) = \{q' \in Q|(\exists q \in Q')q' = \xi(q, \sigma)\}$. 
Let $L(G)$ and $L_{m}(G)$ denote the closed-behavior and the marked behavior, respectively. When $Q_{m} = Q$, we shall also write $G = (Q, \Sigma, \xi, q_{0})$ for simplicity. 
A finite state automaton $G = (Q, \Sigma, \xi, q_{0}, Q_{m})$ is said to be non-blocking if every reachable state in $G$ can reach some marked state in $Q_{m}$ \cite{wonham2015supervisory}.
The ``unobservable reach'' of the state $q \in Q$ under the subset of events $\Sigma' \subseteq \Sigma$ is given by $UR_{G, \Sigma - \Sigma'}(q) := \{q' \in Q|[\exists s \in (\Sigma - \Sigma')^{*}] \, q' = \xi(q,s)\}$.
We shall abuse the notation and define  $P_{\Sigma'}(G)$ to be the finite state automaton $(2^{Q}, \Sigma, \delta, UR_{G, \Sigma - \Sigma'}(q_{0}))$ over $\Sigma$, where $UR_{G, \Sigma - \Sigma'}(q_{0})$ is the initial state, and the (partial) transition function $\delta: 2^{Q} \times \Sigma \rightarrow 2^{Q}$ is defined as follows:

\begin{figure*}[htbp]
\begin{center}
\includegraphics[height=7.7cm]{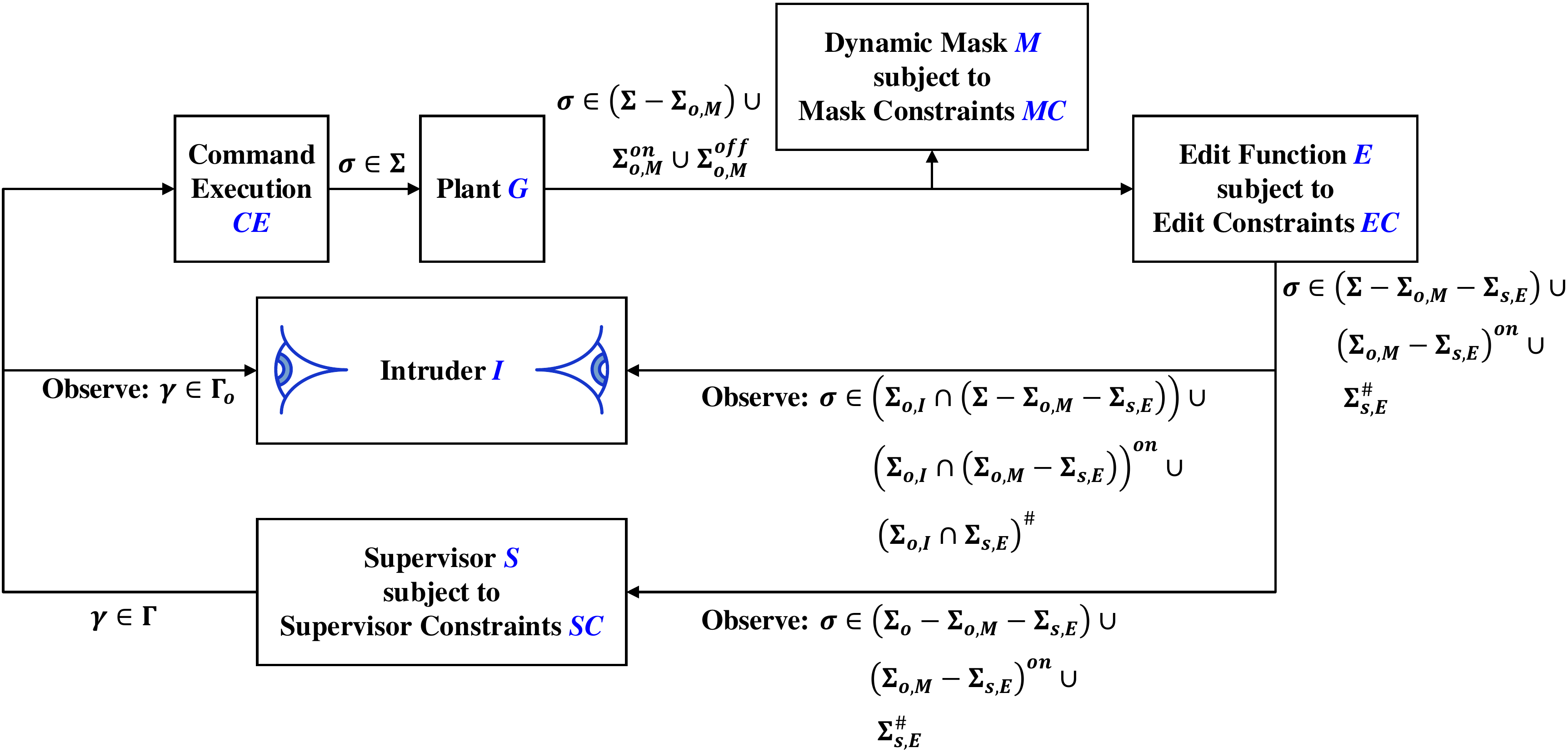}   
\caption{Privacy-preserving control architecture with dynamic mask, edit function and supervisor}
\label{fig:Supervisory control architecture under edit function}
\end{center}        
\end{figure*}

\begin{enumerate}[1)]
    \setlength{\itemsep}{3pt}
    \setlength{\parsep}{0pt}
    \setlength{\parskip}{0pt}
    \item For any $\varnothing \neq Q' \subseteq Q$ and any $\sigma \in \Sigma'$, $\delta(Q', \sigma) = UR_{G, \Sigma - \Sigma'}(\xi(Q', \sigma))$, where
    \[
    UR_{G, \Sigma - \Sigma'}(Q'') = \bigcup\limits_{q \in Q''}UR_{G, \Sigma - \Sigma'}(q)
    \]
    for any $Q'' \subseteq Q$;
    \item For any $\varnothing \neq Q' \subseteq Q$ and $\sigma \in \Sigma - \Sigma'$, $\delta(Q', \sigma) = Q'$.
\end{enumerate}
%We here emphasize that $P_{\Sigma'}(G)$ is over $\Sigma$, instead of $\Sigma'$, and there is no transition defined at the state $\varnothing \in 2^{Q}$.
%As usual, for any two finite state automaton $G_{1} = (Q_{1}, \Sigma_{1}, \xi_{1}, q_{1,0}, Q_{1,m})$ and $G_{2} = (Q_{2}, \Sigma_{2}, \xi_{2}, q_{2,0}, Q_{2,m})$, where $En_{G_{1}}(q) = \{\sigma|\xi_{1}(q, \sigma)!\}$ and $En_{G_{2}}(q) = \{\sigma|\xi_{2}(q, \sigma)!\}$, their synchronous product is denoted as $G_{1}||G_{2} := (Q_{1} \times Q_{2}, \Sigma_{1} \cup \Sigma_{2}, \zeta, (q_{1,0}, q_{2,0}), Q_{1,m} \times Q_{2,m})$, where the (partial) transition function $\zeta$ is defined as follows, for all $(q_{1}, q_{2}) \in Q_{1} \times Q_{2}$, and $\sigma \in \Sigma$:
%\[
%\begin{aligned}
%& \zeta((q_{1}, q_{2}), \sigma) := \\ & \left\{
%\begin{array}{lcl}
%(\xi_{1}(q_{1}, \sigma), \xi_{2}(q_{2}, \sigma))  &      & {\rm if} \, {\sigma \in En_{G_{1}}(q_{1}) \cap En_{G_{2}}(q_{2}),}\\
%(\xi_{1}(q_{1}, \sigma), q_{2})       &      & {\rm if} \, {\sigma \in En_{G_{1}}(q_{1}) \backslash \Sigma_{2},}\\
%(q_{1}, \xi_{2}(q_{2}, \sigma))       &      & {\rm if} \, {\sigma \in En_{G_{2}}(q_{2}) \backslash \Sigma_{1},}\\
%{\rm not \, defined}  &      & {\rm otherwise.}
%\end{array} \right.
%\end{aligned}
%\]
Propositional formulas (or, Boolean formulas) \cite{biere2009handbook} are constructed from (Boolean) variables by using logical connectives ($\wedge,\vee,\neg, \Rightarrow, \Leftrightarrow$). The truth value of a propositional formula $\phi$ is determined by the  truth values of the set $Var(\phi)$  of  variables  which occur in $\phi$. A model of $\phi$ is a map $\mathcal{M}: Var(\phi) \rightarrow \{0,1\}$, where 0 represents false and 1 represents true, such that $\phi$ is evaluated to be true if all the variables $x_i$ in $\phi$ are substituted by $\mathcal{M}(x_i)$. A propositional formula $\phi$ is said to be satisfiable if it has a model $\mathcal{M}$. The Boolean Satisfiability Problem (abbreviated as SAT) is the problem of determining if a given propositional formula is satisfiable.

\textbf{Notation.} Let $\Gamma = 2^{\Sigma_c}-\{\varnothing\}$ denote the set of all control commands, where each control command $\gamma \in \Gamma$ specifies the set of controllable events that are enabled by $\gamma$. This deviates from the standard definition of $\Gamma$. In particular, since uncontrollable events are always allowed to be fired, a control command in this work does not include any uncontrollable event. In this work, it is assumed that when no control command is received by plant $G$, then only uncontrollable events could be executed.
%and $(\Gamma^{*})^{\varnothing} = \Gamma^{*} \cup \{\varnothing\}$. 
%We will often use $\varnothing$ to physically denote ``empty message'' and
For a set $\Sigma$, we use $\Sigma^{on}/\Sigma^{off}/\Sigma^{\#}$ to denote a copy of $\Sigma$ with superscript ``$on$''/``$off$''/``$\#$'' attached to each element in $\Sigma$. 
%Intuitively speaking, the superscript ``$on$'' denotes that the dynamic mask turns on some sensor, ``$off$'' denotes that the dynamic mask turns off some sensor, and ``$\#$'' denotes that the edit function sends some observation event to the supervisor. 
The specific meanings of the relabelled events will be introduced later in Section \ref{sec:Component Models of DES under Mask-Edit Function-Supervisor}.

%%%%%%%%%%%%%%%%%%%%%%%%%%%%%%%%%%%%%%%%%%%%%%%%%%%%%%%%%%%%%%%%%%%%%%%%%%%%%%%%

\section{Component Models with Mask-Edit Function-Supervisor}
\label{sec:Component Models of DES under Mask-Edit Function-Supervisor}

The architecture of privacy-preserving supervisory control system against sensor-actuator eavesdropping intruder via mask, edit function, and supervisor is illustrated in Fig. \ref{fig:Supervisory control architecture under edit function}, where the components are listed as follows:
\begin{itemize}
\setlength{\itemsep}{3pt}
\setlength{\parsep}{0pt}
\setlength{\parskip}{0pt}
    \item Dynamic mask $M$ (subject to mask constraints $MC$).
    \item Edit function $E$ (subject to edit constraints $EC$).
    \item Supervisor $S$ (subject to supervisor constraints $SC$).
    \item Plant $G$.
    \item Command execution component $CE$. 
    \item Intruder $I$.
\end{itemize}
In Fig. \ref{fig:Supervisory control architecture under edit function}, it is assumed that the information flow is as follows:
\begin{enumerate}[1.]
\setlength{\itemsep}{3pt}
\setlength{\parsep}{0pt}
\setlength{\parskip}{0pt}
    \item  Firstly, events fired by $G$ would be observed by the dynamic mask $M$ and edit function $E$, then $M$ would dynamically turn on or turn off sensors depending on its observation and masking capability, which would affect the ability to perceive some sensor events in the future, and $E$ might carry out edit operations and output new information flow to the supervisor $S$.
    \item Then, $S$ and the intruder $I$ would observe the messages which have been altered by $M$ and $E$. $S$ issues control commands to the plant $G$, based on the altered observation messages.
\end{enumerate}
In the following subsections, we shall introduce how to model the above-mentioned six components. 

\subsection{Dynamic mask}
\label{subsec:Dynamic mask}

The dynamic mask can turn on or turn off some sensors when it observes some event.  %fired by the plant $G$. 
In this work, the set of deployed sensors is denoted as $\Delta = \{s_{1}, s_{2}, \dots, s_{n}\}$, where $s_{i}$ is the $i$-th sensor. The set of events that can be sensed by $s_{i}$ is denoted as $\Sigma_{s}^{i} \subseteq \Sigma_{o}$, where $\Sigma_{o}$ is the set of observable events for the supervisor. In practice, it is often the case that $\Sigma_{o}$ is the disjoint unions of $\Sigma_{s}^{i}$'s, i.e., $\Sigma_{o} = \dot{\bigcup}_{i \in \{1,2,\dots,n\}}\Sigma_{s}^{i}$. 
The indices of the sensors that can be masked by the dynamic mask are denoted as $\mathcal{I}_{ms} \subseteq \{1,2,\dots,n\}$, and then the sensors that could be masked is denoted as $\Delta_{m} = \{s_{i} \in \Delta| i \in \mathcal{I}_{ms}\}$. 
For notational convenience, we denote $\Sigma_{o,M} = \dot{\bigcup}_{i \in \mathcal{I}_{ms}}\Sigma_{s}^{i}$, which is the set of events that could be influenced by the mask operation. 
Next, we shall introduce two models: 1) mask constraints, which are utilized to serve as a ``template'' to describe the capabilities of the dynamic mask; 2) dynamic mask, which is one of the desired components that we aim to synthesize. 
%{\color{red} in practice, the sensors are initially open, the dynamic mask can close and open some sensors upon the observation of some event, making the associated events for the sensor unobservable during the time window when the sensor is turned off. This is almost the same as this work's formulation, but with a slight difference. For example, if water level sensor is turned off, then high and level are simultaneously turned off. While this work still allows on one and off the other one. But this is straightforward to model. We may explain this to further convince that dynamic mask cannot be embedded into an edit function due to this simultaneous on-off constraint. }

\begin{figure}[htbp]
\begin{center}
\includegraphics[height=5.2cm]{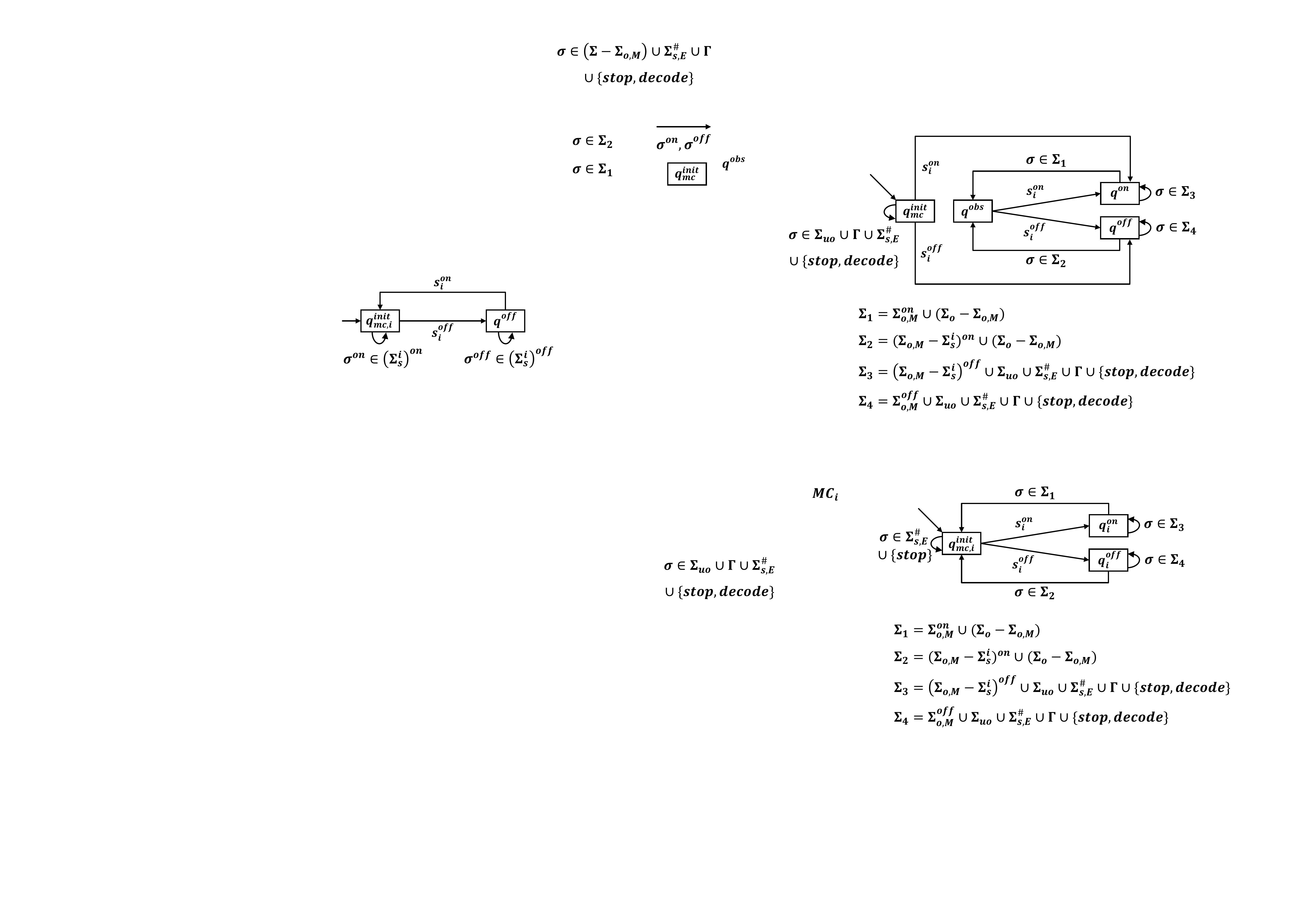}   
\caption{The (schematic) model for mask constraints of $s_{i}$}
\label{fig:The (schematic) model for mask constraints}
\end{center}        
\end{figure}

\textbf{Mask Constraints:} The mask constraints $MC$ is the synchronous product of each sensor $s_{i}$'s mask constraints $MC_{i}$, where $i \in \mathcal{I}_{ms}$. For the sensor $s_{i}$, $MC_{i}$ is modeled as a finite state automaton, which is shown in Fig. \ref{fig:The (schematic) model for mask constraints}:  %{\color{red} comment: $q_{mc,i}^{init}$ and $q^{obs}$ can be merged? If we assume initially the sensor is turned on, then the model can be further simplified. see wechat.. the turn on and turn off events are local to the centralized mask, so other moduled need not model these events. Plant need on and off relabelling. command execution need enable both on and off versions. seems like no other change is necessary.} 
\[
MC_{i} = (Q_{mc,i}, \Sigma_{mc,i}, \xi_{mc,i}, q_{mc,i}^{init}, Q_{mc,i,m})
\]
\begin{itemize}
\setlength{\itemsep}{3pt}
\setlength{\parsep}{0pt}
\setlength{\parskip}{0pt}
    \item $Q_{mc,i} = \{q_{mc,i}^{init}, q_{i}^{on}, q_{i}^{off}\}$
    \item $\Sigma_{mc,i} = (\Sigma - \Sigma_{o,M}) \cup \Sigma_{o,M}^{on} \cup \Sigma_{o,M}^{off} \cup \{s_{i}^{on}, s_{i}^{off}\} \cup \Sigma_{s,E}^{\#} \cup \Gamma \cup \{stop, decode\}$
    \item $\xi_{mc}: Q_{mc,i} \times \Sigma_{mc,i} \rightarrow Q_{mc,i}$
    %\item $Q_{mc,m} = \{q_{mc,i}^{init}\}$
\end{itemize}
The (partial) transition function $\xi_{mc,i}$ is defined as follows:
\begin{enumerate}[1.]
\setlength{\itemsep}{3pt}
\setlength{\parsep}{0pt}
\setlength{\parskip}{0pt}
    \item $\xi_{mc,i}(q_{mc,i}^{init}, s_{i}^{on}) = q_{i}^{on}$ and $\xi_{mc,i}(q_{mc,i}^{init}, s_{i}^{off}) = q_{i}^{off}$
    \item For any $\sigma \in \Sigma_{1} = \Sigma_{o,M}^{on} \cup (\Sigma_{o} - \Sigma_{o,M})$, $\xi_{mc,i}(q_{i}^{on}, \sigma) = q_{mc,i}^{init}$.
    \item For any $\sigma \in \Sigma_{3} = (\Sigma_{o,M} - \Sigma_{s}^{i})^{off} \cup \Sigma_{uo} \cup \Sigma_{s,E}^{\#} \cup \Gamma \cup \{stop, decode\}$, $\xi_{mc,i}(q_{i}^{on}, \sigma) = q_{i}^{on}$.
    \item For any $\sigma \in \Sigma_{2} = (\Sigma_{o,M} - \Sigma_{s}^{i})^{on} \cup (\Sigma_{o} - \Sigma_{o,M})$, $\xi_{mc,i}(q_{i}^{off}, \sigma) = q_{mc,i}^{init}$.
    \item For any $\sigma \in \Sigma_{4} = \Sigma_{o,M}^{off} \cup \Sigma_{uo} \cup \Sigma_{s,E}^{\#} \cup \Gamma \cup \{stop, decode\}$, $\xi_{mc,i}(q_{i}^{off}, \sigma) = q_{i}^{off}$.
    \item For any $\sigma \in \Sigma_{s,E}^{\#} \cup \{stop\}$, $\xi_{mc,i}(q_{mc,i}^{init}, \sigma) = q_{mc,i}^{init}$.
\end{enumerate}
Next, we shall present some explanations for the model $MC_{i}$. For the state set $Q_{mc,i}$, $q_{mc,i}^{init}$ is the initial state denoting that the dynamic mask is ready to turn on or turn off sensor $s_i$ since the system initiation or the last observation of some event. $q_{i}^{on}$ ($q_{i}^{off}$, respectively) denotes the state that the dynamic mask has just turned on (off, respectively) the sensor $s_{i}$.
In the event set, any $\sigma^{on} \in (\Sigma_{s}^{i})^{on}$ is a relabelled copy of the event $\sigma \in \Sigma_{s}^{i}$, denoting an event that can be sensed because of the turned on sensor $s_{i}$. Similarly, any $\sigma^{off} \in (\Sigma_{s}^{i})^{off}$ is a relabelled copy of the event $\sigma \in \Sigma_{s}^{i}$, denoting an event that cannot be sensed because of the turned off sensor $s_{i}$. $s_{i}^{on}$ and $s_{i}^{off}$ denote the events of turning on and turning off the sensor $s_{i}$ by the dynamic mask, respectively, which are controllable and observable to the dynamic mask. Thus, in the following text, we shall denote $\Delta_{m}^{on}$ and $\Delta_{m}^{off}$ as the set of events of turning on and turning off the sensors in $\Delta_{m}$ by the dynamic mask, respectively. Any $\sigma \in \Sigma_{s,E}^{\#} \cup \{stop\}$ denotes the event of edit operations by the edit function, and the event $decode$ denotes that the intruder infers the secret, which shall be introduced later in Section \ref{subsec:Edit function} and \ref{subsec:Intruder}.
For the (partial) transition function $\xi_{mc,i}$, the principle is that once the events $s_{i}^{on}$ ($s_{i}^{off}$, respectively) happens, the state of $MC_{i}$ would transit to state $q_{i}^{on}$ ($q_{i}^{off}$, respectively), denoted by Case 1 in $\xi_{mc,i}$, then, i) only the event in $\Sigma_{1}$ ($\Sigma_{2}$, respectively) might be observed and would enable $MC_{i}$ to transit to the initial state, denoted by Cases 2 and 4, ii) any event in $\Sigma_{3}$ ($\Sigma_{4}$, respectively) is unobservable and would lead to a self-loop, denoted by Cases 3 and 5. Since the edit operations initiated by the edit function could preempt the the mask operations upon the observation of some event, self-loops labelled by events in $\Sigma_{s,E}^{\#} \cup \{stop\}$ are defined at the initial state, denoted by Case 6.

The generated mask constraints $MC=\lVert_{i \in \mathcal{I}_{ms}} MC_i$  is denoted as $MC = (Q_{mc}, \Sigma_{mc}, \xi_{mc}, q_{mc}^{init}, Q_{mc,m})$. Based on the model of $MC$, we have $|Q_{mc}| \leq 3^{|\mathcal{I}_{ms}|}$.

\textbf{Dynamic Mask:} The dynamic mask is modeled as a finite state automaton $M = (Q_{m}, \Sigma_{m}, \xi_{m}, q_{m}^{init}, Q_{m,m})$, where $\Sigma_{m} = (\Sigma - \Sigma_{o,M}) \cup \Sigma_{o,M}^{on} \cup \Sigma_{o,M}^{off} \cup \Delta_{m}^{on} \cup \Delta_{m}^{off} \cup \Sigma_{s,E}^{\#} \cup \Gamma \cup \{stop, decode\}$. 
%Any element in $\Sigma_{s,E}^{\#} \cup \Gamma \cup \{stop, decode\}$ denotes the event happening in the other four components: edit function, supervisor, command execution component, and intruder, which will be introduced in details later in Section \ref{subsec:Edit function}, \ref{subsec:Supervisor}, \ref{subsec:Command execution}, and \ref{subsec:Intruder}. 
For the dynamic mask $M$, the following constraints should be satisfied:
\begin{itemize}
\setlength{\itemsep}{2pt}
\setlength{\parsep}{0pt}
\setlength{\parskip}{0pt}
    \item (M-controllability) For any state $q \in Q_{m}$ and any $\sigma \in \Sigma_{m,uc} = \Sigma_{m} - \Sigma_{m,c} = \Sigma_{m} - (\Delta_{m}^{on} \cup \Delta_{m}^{off})$, $\xi_{m}(q,\sigma)!$.
    \item (M-observability) For any state $q \in Q_{m}$ and any $\sigma \in \Sigma_{m,uo} = \Sigma_{m} - \Sigma_{m,o} = \Sigma_{m} - ((\Sigma_{o} - \Sigma_{o,M}) \cup \Sigma_{o,M}^{on} \cup \Delta_{m}^{on} \cup \Delta_{m}^{off})$, if $\xi_{m}(q,\sigma)!$, then $\xi_{m}(q,\sigma) = q$.
\end{itemize}
M-controllability states that the dynamic mask can only disable events in $\Delta_{m}^{on} \cup \Delta_{m}^{off}$. M-observability states that the dynamic mask can only make a state change after observing an event in $(\Sigma_{o} - \Sigma_{o,M}) \cup \Sigma_{o,M}^{on} \cup \Delta_{m}^{on} \cup \Delta_{m}^{off}$. %\footnote{In the model of the dynamic mask, we shall count $\Sigma_{o,M}^{on} \cup \Sigma_{o,M}^{off}$ in the observable events. The readers could also choose to do not count them in and the methodology proposed in this paper still works.} 
In this work, by construction, all the controllable events for the dynamic mask are also observable to the dynamic mask. In the following text, we shall refer to $\mathcal{C}_{m} = (\Sigma_{m,c}, \Sigma_{m,o})$ as the dynamic mask-control constraint.

\subsection{Edit function}
\label{subsec:Edit function}
The edit function is a component that could instantaneously implement insertion, deletion, and replacement operation when it observes some event. In this work, the basic assumptions of the edit function are given as follows:
\begin{itemize}
    \item The set of observable events for the edit function is denoted as $\Sigma_{o,E} \subseteq \Sigma_{o}$. The set of editable events for the edit function is denoted as $\Sigma_{s,E} \subseteq \Sigma_{o,E}$. %Thus, for any event in $\Sigma_{s,E}$, the edit function could observe it and implement insertion, deletion, and replacement operation on such event. For any event in $\Sigma_{o,E} - \Sigma_{s,E}$, the edit function could observe it but cannot implement edit operations on such event.
    \item The edit function could determine and conduct edit operations when it observes any event in $\Sigma_{o,E}$. When an edit operation is initiated for a specific observation by the edit function, it will be completed before the next observation. For any edit operation after one observation, the output of the edit function is bounded by $U$, i.e., we consider bounded edit function.
    %\item The edit operation initiated by the edit function is instantaneous. 
\end{itemize}
Next, based on the above assumptions, we shall introduce two models: 1) edit constraints, which are utilized to serve as a ``template'' to describe the capabilities of the edit function; 2) edit function, which is one of the desired components that we aim to synthesize.

\begin{figure}[htbp]
\begin{center}
\includegraphics[height=5.9cm]{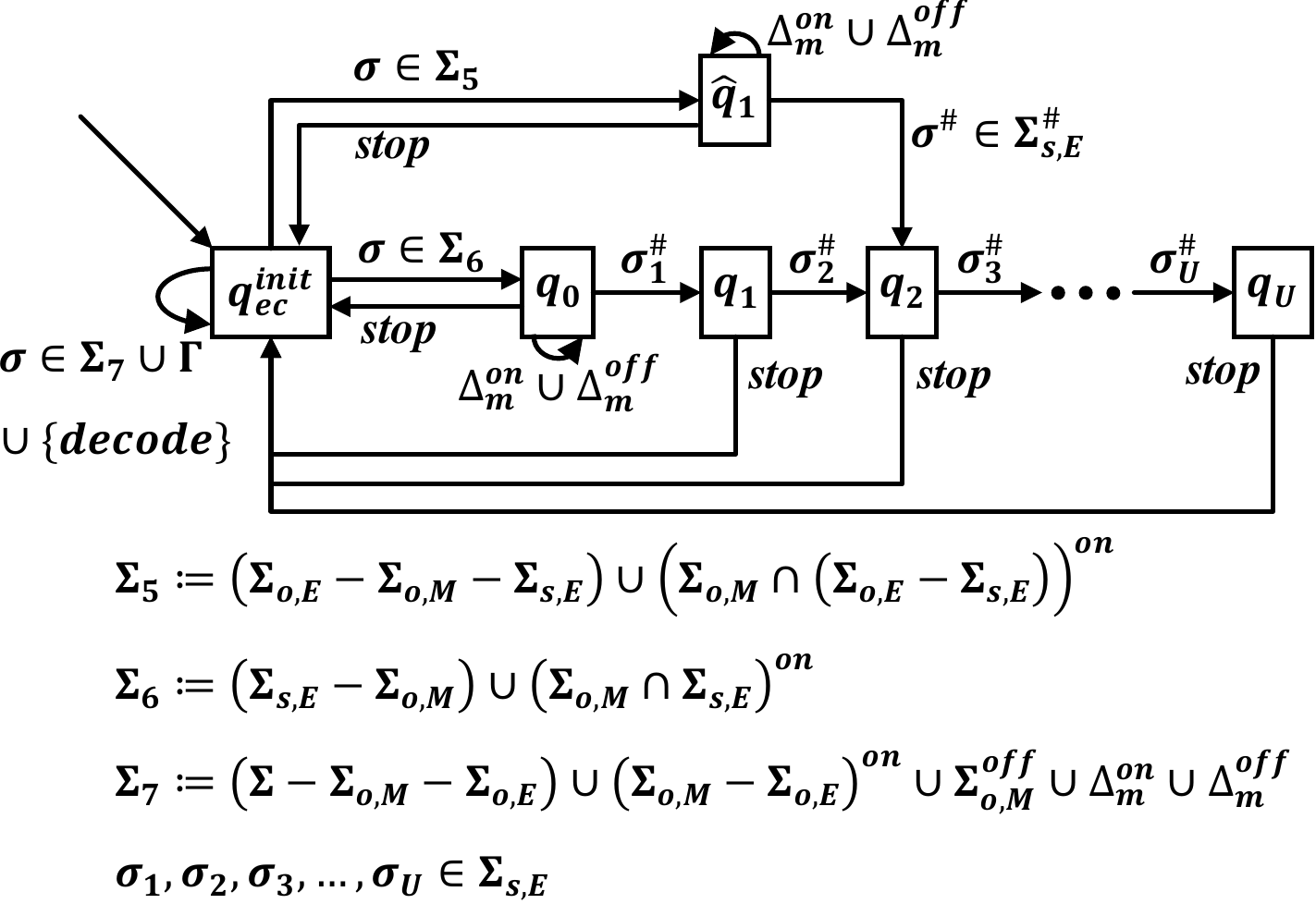}   
\caption{The (schematic) model for edit constraints}
\label{fig:The (schematic) model for editor constraints}
\end{center}        
\end{figure}

\textbf{Edit Constraints:} Firstly, due to the existence of the dynamic mask shown in Fig. \ref{fig:Supervisory control architecture under edit function}, all the events in $\Sigma_{o,M}$ are relabelled as the copies in $\Sigma_{o,M}^{on}$ if the corresponding sensors are not turned off by the dynamic mask. Hence, the set of  events that can be observed by the edit function is $(\Sigma_{o, E}-\Sigma_{o, M}) \cup (\Sigma_{o,M} \cap \Sigma_{o,E})^{on} = \Sigma_{5} \cup \Sigma_{6}$, where $\Sigma_{5}$ and $\Sigma_{6}$ are given in Fig. \ref{fig:The (schematic) model for editor constraints}.
Then, the edit constraints $EC$ is modeled as a finite state automaton, shown in Fig. \ref{fig:The (schematic) model for editor constraints}, similar to the construction in \cite{tai2021privacy}.
\[
EC = (Q_{ec}, \Sigma_{ec}, \xi_{ec}, q_{ec}^{init}, Q_{ec,m})
\]
\begin{itemize}
\setlength{\itemsep}{3pt}
\setlength{\parsep}{0pt}
\setlength{\parskip}{0pt}
    \item $Q_{ec} = \{q_{n}|n \in [0:U]\} \cup \{q_{ec}^{init}, \hat{q}_{1}\}$
    \item $\Sigma_{ec} = (\Sigma - \Sigma_{o,M}) \cup \Sigma_{o,M}^{on} \cup \Sigma_{o,M}^{off} \cup \Delta_{m}^{on} \cup \Delta_{m}^{off} \cup \Sigma_{s,E}^{\#} \cup \Gamma \cup \{stop, decode\}$
    \item $\xi_{ec}: Q_{ec} \times \Sigma_{ec} \rightarrow Q_{ec}$
    \item $Q_{ec,m} = \{q_{ec}^{init}\}$
\end{itemize}
The (partial) transition function $\xi_{ec}$ is defined as follows:
\begin{enumerate}[1.]
\setlength{\itemsep}{3pt}
\setlength{\parsep}{0pt}
\setlength{\parskip}{0pt}
    \item For any $\sigma \in \Sigma_{7} \cup \Gamma \cup \{decode\} = (\Sigma - \Sigma_{o,M} - \Sigma_{o,E}) \cup (\Sigma_{o,M} - \Sigma_{o,E})^{on} \cup \Sigma_{o,M}^{off} \cup \Delta_{m}^{on} \cup \Delta_{m}^{off} \cup \Gamma \cup \{decode\}$, $\xi_{ec}(q_{ec}^{init}, \sigma) = q_{ec}^{init}$.
    \item For any $\sigma \in \Sigma_{6} = (\Sigma_{s,E} - \Sigma_{o,M}) \cup (\Sigma_{o,M} \cap \Sigma_{s,E})^{on}$, $\xi_{ec}(q_{ec}^{init}, \sigma) = q_{0}$.
    \item For any $\sigma \in \Sigma_{5} = (\Sigma_{o,E} - \Sigma_{o,M} - \Sigma_{s,E}) \cup (\Sigma_{o,M} \cap (\Sigma_{o,E} - \Sigma_{s,E}))^{on}$, $\xi_{ec}(q_{ec}^{init}, \sigma) = \hat{q}_{1}$.
    \item For any $n \in [0:U-1]$ and any $\sigma \in \Sigma_{s,E}$, $\xi_{ec}(q_{n}, \sigma^{\#}) = q_{n+1}$ and $\xi_{ec}(\hat{q}_{1}, \sigma^{\#}) = q_{2}$.
    \item For any $q \in \{q_{n}|n \in [0:U]\} \cup \{\hat{q}_{1}\}$, $\xi_{ec}(q, stop) = q_{ec}^{init}$.
    \item For any $\sigma \in \Delta_{m}^{on} \cup \Delta_{m}^{off}$ and any $q \in \{q_{0}, \hat{q}_{1}\}$, $\xi_{ec}(q, \sigma) = q$. 
\end{enumerate}

Next, we shall present some explanations for the model $EC$. For the state set $Q_{ec}$, 1) $q_{ec}^{init}$ is the initial state, where the edit function has not observed any event in $\Sigma_{5} \cup \Sigma_{6}$ since the system initiation or the last edit operation, 2) $q_{n}(n \in [0:U])$ is a state denoting that the edit function has already sent $n$ events since the last observation of some event, 3)$\hat{q}_{1}$ is a state denoting that the edit function has observed some non-editable event in $\Sigma_{5}$, in this case, then we shall count this event in the output of the edit function, i.e., the edit function could still insert at most $U-1$ editable events.
In the event set, any $\sigma^{\#} \in \Sigma_{s,E}^{\#}$ denotes the event of sending an event $\sigma \in \Sigma_{s,E}$ by the edit function, and $stop$ denotes the end of the current round of edit operation.
%any $\sigma^{\#} \in \Sigma_{s,E}^{\#}$ denotes the event of sending $\sigma \in \Sigma_{s,E}$ by the edit function, and the event $stop$ denotes the end of current round of edit operation, which can be controlled and observed by the edit function. 
%Based on the information flow pattern shown in Fig. \ref{fig:Supervisory control architecture under edit function}, since any $\sigma \in \Sigma_{o,M}$ has been relabelled as $\sigma^{on} \in \Sigma_{o,M}^{on}$ if turned on by the dynamic mask, in the model of $EC$, the set of events that could be observed by the edit function is $((\Sigma - \Sigma_{o,M}) \cap \Sigma_{o,E}) \cup (\Sigma_{o,M} \cap \Sigma_{o,E})^{on}$.
All of the events in $\Sigma_{7} \cup \Gamma \cup \{decode\}$ are unobservable and uncontrollable to the edit function.

For the (partial) transition function $\xi_{ec}$,
\begin{itemize}
\setlength{\itemsep}{2pt}
\setlength{\parsep}{0pt}
\setlength{\parskip}{0pt}
    \item Case 1 says that, at the state $q_{ec}^{init}$, if any event $\sigma \in \Sigma_{7} \cup \Gamma \cup \{decode\}$ happens, no edit operation would be implemented since the edit function cannot observe $\sigma$. The occurrence of such event will lead to a self-loop defined at the state $q_{ec}^{init}$. 
    %The purpose of adding case 1 is to ensure 1) the alphabet is $(\Sigma - \Sigma_{o,M}) \cup \Sigma_{o,M}^{on} \cup \Sigma_{o,M}^{off} \cup \Delta_{m}^{on} \cup \Delta_{m}^{off} \cup \Sigma_{s,E}^{\#} \cup \Gamma \cup \{stop, decode\}$, and 2) any event $\sigma \in \Sigma_{3} \cup \Gamma \cup \{decode\}$ is not defined at other states and thus any event in $\Sigma_{1} \cup \Sigma_{2}$ is immediately followed by an event in $\Sigma_{s,E}^{\#} \cup \{stop\}$ to simulate the immediate edit operation or end of the edit operation following the observation of an event in $\Sigma_{1} \cup \Sigma_{2}$.
    \item Case 2 says that, at the state $q_{ec}^{init}$, after the edit function observes any event $\sigma \in (\Sigma_{s,E} - \Sigma_{o,M}) \cup (\Sigma_{o,M} \cap \Sigma_{s,E})^{on}$, it would transit to the state $q_{0}$, at which it could either delete this observed event or replace it with any event in $\Sigma_{s,E}$.
    \item Case 3 says that, at the state $q_{ec}^{init}$, after the edit function observes any event $\sigma \in (\Sigma_{o,E} - \Sigma_{o,M} - \Sigma_{s,E}) \cup (\Sigma_{o,M} \cap (\Sigma_{o,E} - \Sigma_{s,E}))^{on}$, it would transit to the state $\hat{q}_{1}$ and cannot implement any edit operation on this observed event since it is not editable. 
    \item Case 4 says that at any state $q_{n}(n \in [0:U-1])$ and state $\hat{q}_{1}$, the edit function could insert any editable event $\sigma \in \Sigma_{s,E}$. 
    \item Case 5 says that at any state $q_{n}(n \in [0:U])$ and state $\hat{q}_{1}$, the edit function could end the current round of the edit operation and transit back to the initial state $q_{ec}^{init}$.
    \item Case 6 ensures that when the edit function observes some event, either of the edit operation and the mask operation initiated by the dynamic mask may occur first.
\end{itemize}
Based on the model of $EC$, we have $|Q_{ec}| = U + 3$.

\textbf{Edit Function:} The edit function is modeled as a finite state automaton $E = (Q_{e}, \Sigma_{e}, \xi_{e}, q_{e}^{init}, Q_{e,m})$, where $\Sigma_{e} = \Sigma_{ec} = (\Sigma - \Sigma_{o,M}) \cup \Sigma_{o,M}^{on} \cup \Sigma_{o,M}^{off} \cup \Delta_{m}^{on} \cup \Delta_{m}^{off} \cup \Sigma_{s,E}^{\#} \cup \Gamma \cup \{stop, decode\}$, that satisfies the following constraints:
\begin{itemize}
\setlength{\itemsep}{2pt}
\setlength{\parsep}{0pt}
\setlength{\parskip}{0pt}
    \item (E-controllability) For any state $q \in Q_{e}$ and any $\sigma \in \Sigma_{e,uc} = \Sigma_{e} - \Sigma_{e,c} = \Sigma_{e} - (\Sigma_{s,E}^{\#} \cup \{stop\})$, $\xi_{e}(q,\sigma)!$.
    \item (E-observability) For any state $q \in Q_{e}$ and any $\sigma \in \Sigma_{e,uo} = \Sigma_{e} - \Sigma_{e,o} = \Sigma_{e} - ((\Sigma_{o,E} - \Sigma_{o,M}) \cup (\Sigma_{o,M} \cap \Sigma_{o,E})^{on} \cup \Sigma_{s,E}^{\#} \cup \{stop\})$, if $\xi_{e}(q,\sigma)!$, then $\xi_{e}(q,\sigma) = q$.
\end{itemize}
E-controllability states that the edit function can only disable events in $\Sigma_{s,E}^{\#} \cup \{stop\}$. E-observability states that the edit function can only make a state change after observing an event in $(\Sigma_{o,E} - \Sigma_{o,M}) \cup (\Sigma_{o,M} \cap \Sigma_{o,E})^{on} \cup \Sigma_{s,E}^{\#} \cup \{stop\}$. In this work, by construction, all the controllable events for the edit function are also observable to the edit function. In the following text, we shall refer to $\mathcal{C}_{e} = (\Sigma_{e,c}, \Sigma_{e,o})$ as the edit function-control constraint.

\subsection{Supervisor}
\label{subsec:Supervisor}
In this part, we shall introduce two models: 1) supervisor constraints, which are utilized to serve as a ``template'' to describe the capabilities of the supervisor; 2) supervisor, which is one of the desired components that we aim to synthesize.

\begin{figure}[htbp]
\begin{center}
\includegraphics[height=2.7cm]{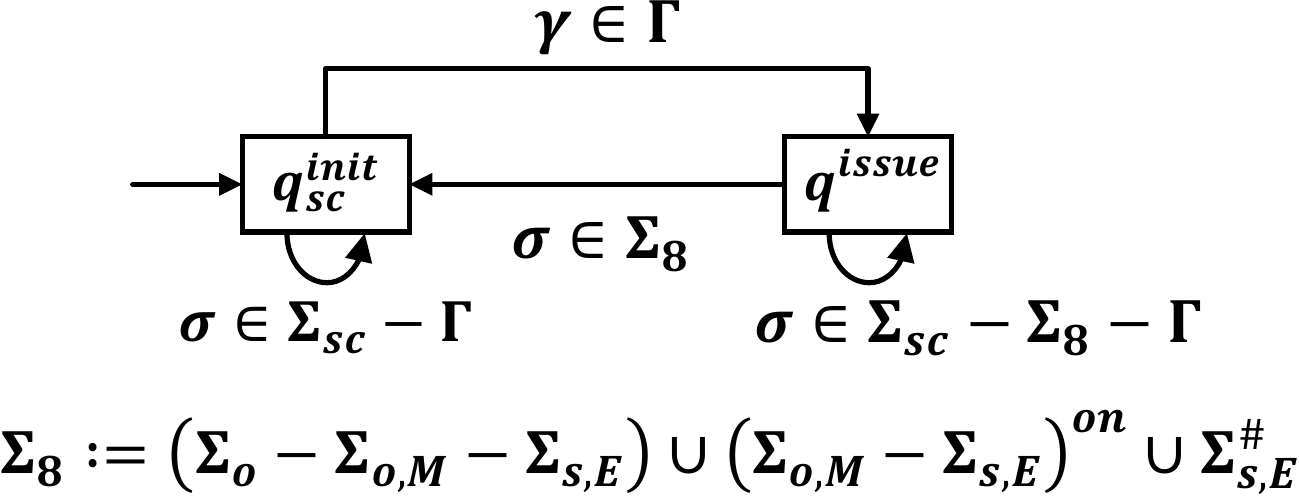}   
\caption{The (schematic) model for supervisor constraints}
\label{fig:The (schematic) model for supervisor constraints}
\end{center}        
\end{figure}

\textbf{Supervisor Constraints:} Firstly, due to the existence of the edit function and the dynamic mask shown in Fig. \ref{fig:Supervisory control architecture under edit function}, all the events in $\Sigma_{s,E}$ have been relabelled as the copies in $\Sigma_{s,E}^{\#}$ and all the events in $\Sigma_{o,M} - \Sigma_{s,E}$ have been relabelled as the copies in $(\Sigma_{o,M} - \Sigma_{s,E})^{on}$. Hence, the set of events that can be observed by the supervisor is denoted as $\Sigma_{8} = (\Sigma_{o} - \Sigma_{o,M} - \Sigma_{s,E}) \cup (\Sigma_{o,M} - \Sigma_{s,E})^{on} \cup \Sigma_{s,E}^{\#}$.
Then, the supervisor constraints are modeled as a finite state automaton shown in Fig. \ref{fig:The (schematic) model for supervisor constraints}. 
Intuitively speaking, when the system initiates, the supervisor could issue the initial control command without observing any event. Afterwards, the supervisor could issue a new control command again only after it has observed at least one  event. 
\[
SC = (Q_{sc}, \Sigma_{sc}, \xi_{sc}, q_{sc}^{init})
\]
\begin{itemize}
\setlength{\itemsep}{3pt}
\setlength{\parsep}{0pt}
\setlength{\parskip}{0pt}
    \item $Q_{sc} = \{q_{sc}^{init}, q^{issue}\}$
    \item $\Sigma_{sc} = (\Sigma - \Sigma_{o,M}) \cup \Sigma_{o,M}^{on} \cup \Sigma_{o,M}^{off} \cup \Delta_{m}^{on} \cup \Delta_{m}^{off} \cup \Sigma_{s,E}^{\#} \cup \Gamma \cup \{stop, decode\}$
    \item $\xi_{sc}: Q_{sc} \times \Sigma_{sc} \rightarrow Q_{sc}$
\end{itemize}
The (partial) transition function $\xi_{sc}$ is defined as follows:
\begin{enumerate}[1.]
\setlength{\itemsep}{3pt}
\setlength{\parsep}{0pt}
\setlength{\parskip}{0pt}
    \item For any $\sigma \in \Sigma_{sc} - \Gamma = (\Sigma - \Sigma_{o,M}) \cup \Sigma_{o,M}^{on} \cup \Sigma_{o,M}^{off} \cup \Delta_{m}^{on} \cup \Delta_{m}^{off} \cup \Sigma_{s,E}^{\#} \cup \{stop, decode\}$, $\xi_{sc}(q_{sc}^{init}, \sigma) = q_{sc}^{init}$.
    \item For any $\gamma \in \Gamma$, $\xi_{sc}(q_{sc}^{init}, \gamma) = q^{issue}$.
    \item For any $\sigma \in \Sigma_{8} = (\Sigma_{o} - \Sigma_{o,M} - \Sigma_{s,E}) \cup (\Sigma_{o,M} - \Sigma_{s,E})^{on} \cup \Sigma_{s,E}^{\#}$, $\xi_{sc}(q^{issue}, \sigma) = q_{sc}^{init}$.
    \item For any $\sigma \in \Sigma_{sc} - \Sigma_{8} - \Gamma = \Sigma_{uo} \cup (\Sigma_{s,E} - \Sigma_{o,M}) \cup (\Sigma_{o,M} \cap \Sigma_{s,E})^{on} \cup \Sigma_{o,M}^{off} \cup \Delta_{m}^{on} \cup \Delta_{m}^{off} \cup \{stop, decode\}$, $\xi_{sc}(q^{issue}, \sigma) = q^{issue}$.
\end{enumerate}
Next, we shall present some explanations for the model $SC$. For the state set $Q_{sc}$, 1) $q_{sc}^{init}$ is the initial state, denoting that the supervisor has not issued any control command since the system initiation or the last observation of some event in $(\Sigma_{o} - \Sigma_{o,M} - \Sigma_{s,E}) \cup (\Sigma_{o,M} - \Sigma_{s,E})^{on} \cup \Sigma_{s,E}^{\#}$, 2) $q^{issue}$ is a state denoting that the supervisor has just issued a control command and since then it has not observed any event in $(\Sigma_{o} - \Sigma_{o,M} - \Sigma_{s,E}) \cup (\Sigma_{o,M} - \Sigma_{s,E})^{on} \cup \Sigma_{s,E}^{\#}$.
In the event set, any $\gamma \in \Gamma$ denotes the event of issuing a control command $\gamma$ by the supervisor. In this work, we shall impose a natural assumption that any $\gamma \in \Gamma$ is observable to the supervisor.

For the (partial) transition function $\xi_{sc}$, 
\begin{itemize}
\setlength{\itemsep}{3pt}
\setlength{\parsep}{0pt}
\setlength{\parskip}{0pt}
    \item Cases 1 and 2 say that, at the state $q_{sc}^{init}$, the supervisor would make a transition to state $q^{issue}$ only after it issues a control command $\gamma \in \Gamma$. If any other event $\sigma \in \Sigma_{sc} - \Gamma$ happens, the supervisor would only do a self-loop transition.
    \item Cases 3 and 4 say that, at the state $q^{issue}$, since the supervisor has just issued a control command, it would not issue a control command again until receiving a new observation. Thus, at the state $q^{issue}$, the supervisor would make a transition to state $q_{sc}^{init}$ only after it observes an event in $(\Sigma_{o} - \Sigma_{o,M} - \Sigma_{s,E}) \cup (\Sigma_{o,M} - \Sigma_{s,E})^{on} \cup \Sigma_{s,E}^{\#}$. The occurrence of any other event in $\Sigma_{sc}$ would lead to a self-loop.
\end{itemize}
Based on the model of $SC$, we have that $|Q_{sc}| = 2$.

\textbf{Supervisor:} The supervisor is modeled as a finite state automaton $S = (Q_{s}, \Sigma_{s}, \xi_{s}, q_{s}^{init}, Q_{s,m})$, where $\Sigma_{s} = \Sigma_{sc} = (\Sigma - \Sigma_{o,M}) \cup \Sigma_{o,M}^{on} \cup \Sigma_{o,M}^{off} \cup \Delta_{m}^{on} \cup \Delta_{m}^{off} \cup \Sigma_{s,E}^{\#} \cup \Gamma \cup \{stop, decode\}$, that satisfies the following constraints:
\begin{itemize}
\setlength{\itemsep}{3pt}
\setlength{\parsep}{0pt}
\setlength{\parskip}{0pt}
    \item (S-controllability) For any state $q \in Q_{s}$ and any $\sigma \in \Sigma_{s} - \Sigma_{s,c} = \Sigma_{s} - \Gamma$, $\xi_{s}(q, \sigma)!$.
    \item (S-observability) For any state $q \in Q_{s}$ and any $\sigma \in \Sigma_{s} - \Sigma_{s,o} = \Sigma_{s} -( (\Sigma_{o} - \Sigma_{o,M} - \Sigma_{s,E}) \cup (\Sigma_{o,M} - \Sigma_{s,E})^{on} \cup \Sigma_{s,E}^{\#} \cup \Gamma)$, if $\xi_{s}(q, \sigma)!$, then $\xi_{s}(q, \sigma) = q$.
\end{itemize}
S-controllability states that the supervisor can only disable events in $\Gamma$. S-observability states that the supervisor can only make a state change after observing events in $(\Sigma_{o} - \Sigma_{o,M} - \Sigma_{s,E}) \cup (\Sigma_{o,M} - \Sigma_{s,E})^{on} \cup \Sigma_{s,E}^{\#} \cup \Gamma$. In this work, by construction, all the controllable events for the supervisor are also observable to the supervisor. In the following text, we shall refer to $\mathcal{C}_{s} = (\Sigma_{s,c}, \Sigma_{s,o})$ as the supervisor-control constraint.

\subsection{Plant}
\label{subsec:Plant}
Plant $G$ is modeled as a finite state automaton $G = (Q, \Sigma, \xi_{G}, q^{init}, Q_{G,m})$. The set of secret states in plant $G$ is denoted as $Q_{sec} \subseteq Q$. The set of bad states that need to be avoided in $G$ is denoted as $Q_{avoid} \subseteq Q$.
In this work, we consider current-state opacity (CSO) \cite{Saboori2007SBO}.

Next, we shall perform a relabelling of $G$ to capture the effects of the dynamic mask turning on and turning off some sensors. Since all the events in $\Sigma_{o,M}$ would be affected by the action of turning on and turning off sensors by the dynamic mask, we shall replace all the transitions labelled by events $\sigma \in \Sigma_{o,M}$ in $G$ with two relabelled copies $\sigma^{on} \in \Sigma_{o,M}^{on}$ and $\sigma^{off} \in \Sigma_{o,M}^{off}$. The modified plant $G_{new}$ is defined as follows:
\[
G_{new} = (Q, (\Sigma - \Sigma_{o,M}) \cup \Sigma_{o,M}^{on} \cup \Sigma_{o,M}^{off}, \xi_{new}, q^{init}, Q_{G,m})
\]
where $\xi_{new}$ is defined as: 1) for any $q, q' \in Q$ and any $\sigma \in \Sigma_{o,M}$, $\xi_{G}(q, \sigma) = q' \Leftrightarrow \xi_{new}(q, \sigma^{on}) = q' \wedge \xi_{new}(q, \sigma^{off}) = q'$, 2) for any $q, q' \in Q$ and any $\sigma \in \Sigma - \Sigma_{o,M}$, $\xi_{G}(q, \sigma) = q' \Leftrightarrow \xi_{new}(q, \sigma) = q'$. For notational simplicity, in the following text, the modified plant is still denoted as $G$, that is, $G := G_{new} = (Q, (\Sigma - \Sigma_{o,M}) \cup \Sigma_{o,M}^{on} \cup \Sigma_{o,M}^{off}, \xi_{G}, q^{init}, Q_{G,m})$.

%\textbf{Definition 1 (CSO)} $G = (Q, \Sigma, \xi_{G}, q^{init}, Q_{G,m})$ is CSO w.r.t. the projection $P$ and the set of secret states $Q_{sec} \subseteq Q$ if
%\[
%\begin{aligned}
%& \forall t \in L_{S} := \{t \in L(G)|\xi_{G}(q^{init}, t) \in Q_{sec}\}, \exists t' \in L_{NS} \\ & := \{t \in L(G)|\xi_{G}(q^{init}, t) \in (Q \backslash Q_{sec})\}, P(t) = P(t')
%\end{aligned}
%\]

\subsection{Command execution component}
\label{subsec:Command execution}
%Firstly, we shall explain why we need to introduce and model this component: As we know, the input to the plant is the control commands in $\Gamma$, while the output of the plant is the events in $\Sigma$. There is thus a need for modelling the ``transduction" from the input $\gamma \in \Gamma$ to the output $\sigma \in \Sigma$. This requires an automaton model over the event set $\Sigma \cup \Gamma$ that models how the control commands are executed, which is referred to as the command execution automaton $CE$.

\begin{figure}[htbp]
\begin{center}
\includegraphics[height=2.6cm]{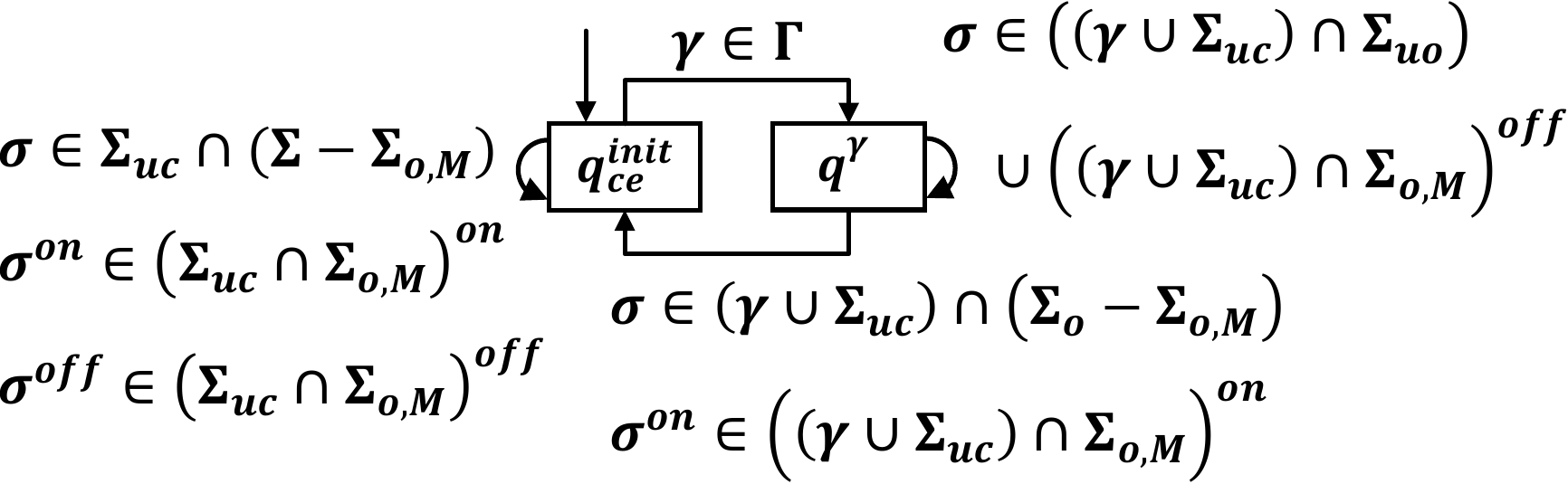}   
\caption{The (schematic) model for command execution component}
\label{fig:The (schematic) model for command execution}
\end{center}        
\end{figure}

The command execution automaton is a component to explicitly describe the execution phase of a control command. Due to the effects of turning on or turning off sensors by the dynamic mask, all the events in $\Sigma_{o,M}$ have been relabelled as the copies in $\Sigma_{o,M}^{on}$ and $\Sigma_{o,M}^{off}$. Then, the command execution automaton is modeled as a finite state automaton $CE$, which is illustrated in Fig. \ref{fig:The (schematic) model for command execution}.
\[
CE = (Q_{ce}, \Sigma_{ce}, \xi_{ce}, q_{ce}^{init}, Q_{ce,m})
\]
\begin{itemize}
\setlength{\itemsep}{3pt}
\setlength{\parsep}{0pt}
\setlength{\parskip}{0pt}
    \item $Q_{ce} = \{q^{\gamma}|\gamma \in \Gamma\} \cup \{q_{ce}^{init}\}$ 
    \item $\Sigma_{ce} = \Gamma \cup (\Sigma - \Sigma_{o,M}) \cup \Sigma_{o,M}^{on} \cup \Sigma_{o,M}^{off}$
    \item $\xi_{ce}: Q_{ce} \times \Sigma_{ce} \rightarrow Q_{ce}$
    \item $Q_{ce,m} = \{q_{ce}^{init}\}$
\end{itemize}
The (partial) transition function $\xi_{ce}$ is defined as follows:
\begin{enumerate}[1.]
\setlength{\itemsep}{3pt}
\setlength{\parsep}{0pt}
\setlength{\parskip}{0pt}
    \item For any $\gamma \in \Gamma$, $\xi_{ce}(q_{ce}^{init}, \gamma) = q^{\gamma}$.
    \item For any $\sigma \in ((\gamma \cup \Sigma_{uc}) \cap \Sigma_{uo}) \cup ((\gamma \cup \Sigma_{uc}) \cap \Sigma_{o,M})^{off}$, $\xi_{ce}(q^{\gamma}, \sigma) = q^{\gamma}$.
    \item For any $\sigma \in (\gamma \cup \Sigma_{uc}) \cap (\Sigma_{o} - \Sigma_{o,M})$, $\xi_{ce}(q^{\gamma}, \sigma) = q_{ce}^{init}$.
    \item For any $\sigma \in (\gamma \cup \Sigma_{uc}) \cap \Sigma_{o,M}$, $\xi_{ce}(q^{\gamma}, \sigma^{on}) = q_{ce}^{init}$.
    \item For any $\sigma \in \Sigma_{uc} \cap (\Sigma - \Sigma_{o,M})$, $\xi_{ce}(q_{ce}^{init}, \sigma) = q_{ce}^{init}$.
    \item For any $\sigma \in \Sigma_{uc}\cap \Sigma_{o,M}$, $\xi_{ce}(q_{ce}^{init}, \sigma^{on}) = q_{ce}^{init}$ and $\xi_{ce}(q_{ce}^{init}, \sigma^{off}) = q_{ce}^{init}$.
\end{enumerate}
Next, we shall present some explanations for the model $CE$. For the state set $Q_{ce}$, 1) $q_{ce}^{init}$ is a state denoting that the command execution automaton is not using any control command, 2) $q^{\gamma} \in Q_{ce} (\gamma \in \Gamma)$ is a state denoting that command execution automaton is using the control command $\gamma$.

For the (partial) transition function $\xi_{ce}$,
\begin{itemize}
\setlength{\itemsep}{3pt}
\setlength{\parsep}{0pt}
\setlength{\parskip}{0pt}
    \item Case 1 says that, at the state $q_{ce}^{init}$, if the command execution component receives a control command $\gamma$, then it will start to use $\gamma$ and transit to the state $q^{\gamma}$.
    \item Case 2 says that, at the state $q^{\gamma}$, if any unobservable event $\sigma \in ((\gamma \cup \Sigma_{uc}) \cap \Sigma_{uo}) \cup ((\gamma \cup \Sigma_{uc}) \cap \Sigma_{o,M})^{off}$ is executed, then it will lead to a self-loop, meaning that $\gamma$ will be reused.
    \item Cases 3 and 4 say that, at the state $q^{\gamma}$, if any event $\sigma \in (\gamma \cup \Sigma_{uc}) \cap (\Sigma_{o} - \Sigma_{o,M})$, or $\sigma^{on} \in ((\gamma \cup \Sigma_{uc}) \cap \Sigma_{o,M})^{on}$ is executed by the command execution component, then it will transit back to the state $q_{ce}^{init}$ and wait for the next control command issued by the supervisor.   
    %{\color{red} when $\sigma^{off}$ is executed, the supervisor cannot observe it. We may need to think about the reason/justify why the supervisor can issue a control command based on $\sigma^{off}$ or explain how the alternative can be modelled easily. Technically, if the sensor is switched off, it is treated as not present and thus the event becomes unobservable, maybe treatment is more similar to unobservable selfloop. But treating sigma off as unobservable may lead to lack of control commands to drive the  system execute}
    \item Cases 5 and 6 say that, at the state $q_{ce}^{init}$, any uncontrollable event $\sigma \in \Sigma_{uc} \cap (\Sigma - \Sigma_{o,M})$, or $\sigma^{on} \in (\Sigma_{uc} \cap \Sigma_{o,M})^{on}$, or $\sigma^{off} \in (\Sigma_{uc} \cap \Sigma_{o,M})^{off}$ can be executed. 
\end{itemize}
Based on the model of $CE$, we have that $|Q_{ce}| = 2^{|\Sigma_{c}|}$.

\subsection{Intruder}
\label{subsec:Intruder}
In Fig. \ref{fig:Supervisory control architecture under edit function}, the intruder is a component that could eavesdrop (part of) the sensor events and (part of) the control commands. 
%Based on its observation, the intruder would take use of such information to carry out state estimation and infer whether the plant $G$ has reached the secret state. 
The assumptions for the intruder in this work are listed as follows: 
\begin{itemize}
\setlength{\itemsep}{3pt}
\setlength{\parsep}{0pt}
\setlength{\parskip}{0pt}
    \item The set of observable sensor events for the intruder is denoted as $\Sigma_{o,I} \subseteq \Sigma$. Here, $\Sigma_{o,I}$ could be different from $\Sigma_{o,M}$, $\Sigma_{o,E}$, and $\Sigma_{o}$. The set of observable control commands for the intruder is denoted as $\Gamma_{o} \subseteq \Gamma$.   
    \item The intruder has the full knowledge of the structure of the plant but does not know the specification of the plant. The intruder knows the controllable and observable events.
    %to construct $CE$ {\color{red} I need to check one problem: to construct $CE$ in this model, does the intruder need to know what is maskable; or we just say "The intruder knows the controllable and observable events", as basically the intruder only needs to know the original model of $CE$. That is, we just relabel the intruder model with $on$ and $sharp$ to perform the synchronization}. 
    In addition, the intruder is not sure about the presence of the dynamic mask and the edit function.
\end{itemize}
Since the intruder has the prior knowledge of the plant and could observe some sensor events and control commands, it is able to compare its online observations with the prior knowledge to check whether there exist some privacy-preserving components in the closed-loop system, that is, it would discover the existence of the dynamic mask or the edit function once some information inconsistency happens. 
Next, we shall introduce how to model the intruder. Generally speaking, the construction of the intruder model consists of the following steps:

\textbf{Step 1:}  Firstly, since the intruder is not sure about the presence of the dynamic mask and the edit function, we shall remove all the transitions labelled by events in $\Sigma_{o,M}^{off}$ in $G$ and $CE$, and generate a modified plant $G_{1} = (Q, \Sigma_{G_{1}}, \xi_{G_{1}}, q^{init}, Q_{G, m})$, where $\Sigma_{G_{1}} = (\Sigma - \Sigma_{o,M}) \cup \Sigma_{o,M}^{on}$, and a modified command execution component $CE_{1} = (Q_{ce}, \Sigma_{ce_{1}}, \xi_{ce_{1}}, q_{ce}^{init}, Q_{ce, m})$, where $\Sigma_{ce_{1}} = (\Sigma - \Sigma_{o,M}) \cup \Sigma_{o,M}^{on} \cup \Gamma$.
Since the intruder could observe (part of) the control commands, each time when it observes one, based on the structure of the plant $G_{1}$ and the command execution component $CE_1$, the intruder is able to refine its state estimation and predict the events that would be fired by the plant. 
To encode this feature, we compute the synchronous product of the command execution component $CE_{1}$, which describes the phase from using a control command to executing an event, and the model of plant $G_{1}$.  
Then, we mark all the states. For notational simplicity, we still denote the resulting automaton, given as follows, as $G_{1}||CE_{1}$:
\[
G_{1}||CE_{1} = (Q_{G_{1}||CE_{1}}, \Sigma_{G_{1}||CE_{1}}, \xi_{G_{1}||CE_{1}}, q_{G_{1}||CE_{1}}^{init})
\]
where $\Sigma_{G_{1}||CE_{1}} = (\Sigma - \Sigma_{o,M}) \cup \Sigma_{o,M}^{on} \cup \Gamma$. In $G_{1}||CE_{1}$, the set of secret states is denoted as $Q_{sec}^{p} = \{(q, q_{ce}) \in Q_{G_{1}||CE_{1}}|q \in Q_{sec}\}$ since  whether a state is secret or not in $G_{1}||CE_{1}$ entirely depends on whether the plant state belongs to $Q_{sec}$.

\textbf{Step 2:} At this step, we need to encode the following features of the intruder: 1) In $G_{1}||CE_{1}$, the intruder could only observe elements in $\widetilde{\Sigma_{o,I}} = (\Sigma_{o,I} - \Sigma_{o,M}) \cup (\Sigma_{o,I} \cap \Sigma_{o,M})^{on} \cup \Gamma_{o}$; 2) Based on the prior knowledge $G_{1}||CE_{1}$, any information inconsistency would be discovered by the intruder based on its partial observation. Thus, we shall construct the following finite state automaton
\[
P_{\widetilde{\Sigma_{o,I}}}(G_{1}||CE_{1}) = (Q_{temp}, \Sigma_{temp}, \xi_{temp}, q_{temp}^{init})
\]
\begin{itemize}
\setlength{\itemsep}{3pt}
\setlength{\parsep}{0pt}
\setlength{\parskip}{0pt}
    \item $Q_{temp} = 2^{Q_{G_{1}||CE_{1}}}= 2^{Q \times Q_{ce}}$ 
    \item $\Sigma_{temp} = (\Sigma - \Sigma_{o,M}) \cup \Sigma_{o,M}^{on} \cup \Gamma$
    \item $\xi_{temp}: Q_{temp} \times \Sigma_{temp} \rightarrow Q_{temp}$
    \item $q_{temp}^{init} = UR_{G, \Sigma_{temp} - \widetilde{\Sigma_{o,I}}}(q_{G_{1}||CE_{1}}^{init})$ 
\end{itemize}
Based on the above-constructed automaton, it is noteworthy that: 1) reaching any state in $2^{Q_{sec}^{p}} - \{\varnothing\}$ implies that the intruder has inferred the system secret %since $P_{\widetilde{\Sigma_{o,I}}}(G_{1}||CE_{1})$ is indeed a state estimator
; 2) any transition to state $\varnothing$ in $P_{\widetilde{\Sigma_{o,I}}}(G_{1}||CE_{1})$ reveals the 
existence of the privacy-preserving components.

\textbf{Step 3:} At this step, for $P_{\widetilde{\Sigma_{o,I}}}(G_{1}||CE_{1})$, we shall explicitly encode the fact that the intruder has inferred the system secret at any state in $2^{Q_{sec}^{p}} - \{\varnothing\}$.
In addition, due to the existence of the edit function, any event $\sigma \in (\Sigma - \Sigma_{o,M}) \cap \Sigma_{s,E} = \Sigma_{s,E} - \Sigma_{o,M}$ and $\sigma^{on} \in (\Sigma_{o,M} \cap \Sigma_{s,E})^{on}$ need to be replaced with the relabelled copy $ \sigma^{\#} \in \Sigma_{s,E}^{\#}$. 
Then, based on $P_{\widetilde{\Sigma_{o,I}}}(G_{1}||CE_{1})$, the model of the intruder $I$ is generated by the following procedure:
\[
I = (Q_{i}, \Sigma_{i}, \xi_{i}, q_{i}^{init}, Q_{i,m})
\]
\begin{enumerate}[1.]
\setlength{\itemsep}{3pt}
\setlength{\parsep}{0pt}
\setlength{\parskip}{0pt}
    \item $Q_{i} = Q_{temp} \cup \{q^{unsafe}\}$
    \item $\Sigma_{i} = (\Sigma - \Sigma_{o,M} - \Sigma_{s,E}) \cup (\Sigma_{o,M} - \Sigma_{s,E})^{on} \cup \Sigma_{s,E}^{\#} \cup \Gamma \cup \{decode\}$
    \item \begin{enumerate}[a.]
    \setlength{\itemsep}{3pt}
    \setlength{\parsep}{0pt}
    \setlength{\parskip}{0pt}
        \item $(\forall q \in Q_{i}) \, q \in 2^{Q_{sec}^{p}} - \{\varnothing\} \Leftrightarrow \xi_{i}(q, decode) = q^{unsafe}$
        \item $(\forall q, q' \in Q_{i})(\forall \sigma \in \Sigma_{s,E} - \Sigma_{o,M}) \, \xi_{temp}(q, \sigma) = q' \Leftrightarrow \xi_{i}(q, \sigma^{\#}) = q'$
        \item $(\forall q, q' \in Q_{i})(\forall \sigma \in \Sigma_{o,M} \cap \Sigma_{s,E}) \, \xi_{temp}(q, \sigma^{on}) = q' \Leftrightarrow \xi_{i}(q, \sigma^{\#}) = q'$
        \item $(\forall q, q' \in Q_{i})(\forall \sigma \in (\Sigma - \Sigma_{o,M} - \Sigma_{s,E}) \cup (\Sigma_{o,M} - \Sigma_{s,E})^{on} \cup \Gamma) \, \xi_{temp}(q, \sigma) = q' \Leftrightarrow \xi_{i}(q, \sigma) = q'$
        \item $(\forall q \in \{\varnothing, q^{unsafe}\})(\forall \sigma \in \Sigma_{i} - \{decode\})\, \xi_{i}(q,\sigma)\\ = q$
    \end{enumerate}
    \item $q_{i}^{init} = q_{temp}^{init}$
    \item $Q_{i,m} = Q_{temp} - \{\varnothing\}$
\end{enumerate}
We shall give some explanations for the above procedure. 
\begin{itemize}
\setlength{\itemsep}{3pt}
\setlength{\parsep}{0pt}
\setlength{\parskip}{0pt}
    \item At Step 1 and Step 2, a new state $q^{unsafe}$ and a new event $decode$, denoting that the intruder infers that plant $G$ has reached a secret state, is added to the state set and the event set, respectively. This generates the new state set $Q_{i}$ and the new event set $\Sigma_{i}$, respectively.
    \item Step 3 is dedicated to defining the (partial) transition functions $\xi_{i}$. At Step 3.a, for any state $q \in 2^{Q_{sec}^{p}} - \{\varnothing\}$, the intruder infers that plant $G$ has reached a secret state; thus, a new transition is added to the state $q$, denoted by $\xi_{i}(q, decode) = q^{unsafe}$. Since the event $decode$ is uncontrollable to the dynamic mask, edit function, and supervisor, it should be avoided that the intruder transits to any state $q \in 2^{Q_{sec}^{p}} - \{\varnothing\}$. 
    At Steps 3.b, 3.c. and 3.d,  transitions are relabelled as discussed before.
    At Step 3.e, for any state $q \in \{\varnothing, q^{unsafe}\}$, where the intruder has either already inferred the system secret or the existence of the privacy-preserving components, self-loops labelled by events in $\Sigma_{i} - \{decode\}$ are defined, as the occurrence of any further event would not change this fact.
\end{itemize}
From the point of view of the dynamic mask, the edit function, and the supervisor, they should work together such that the transitions to the state $\varnothing$ and $q^{unsafe}$ are avoided. Based on the model of $I$, we have that $|Q_{i}| \leq 2^{|Q| \times |Q_{ce}|} + 1$.

\section{Co-Synthesis of Dynamic Mask, Edit Function, and Supervisor}
\label{sec:Co-Synthesis of Dynamic Mask, Edit Function, and Supervisor for Opacity Enforcement}
In this section, based on the component models introduced in Section \ref{sec:Component Models of DES under Mask-Edit Function-Supervisor}, we shall propose an incremental heuristic approach to co-synthesize a dynamic mask, an edit function, and a supervisor for opacity enforcement and requirement satisfaction.

\subsection{Solution Methodology}
\label{subsec:Solution Methodology}
Based on the architecture shown in Fig. \ref{fig:Supervisory control architecture under edit function}, the closed-loop system behavior $\mathcal{B}$ can be modeled as the synchronous product of the plant $G$, command execution component $CE$, dynamic mask constraints $MC$, edit constrains $EC$, supervisor constraints $SC$, intruder $I$, dynamic mask $M$, edit function $E$, and supervisor $S$, which is given as follows:
\[
\begin{aligned}
\mathcal{B} & = G||CE||MC||EC||SC||I||M||E||S \\ & = (Q_{b}, \Sigma_{b}, \xi_{b}, q_{b}^{init}, Q_{b,m})
\end{aligned}
\]
\begin{itemize}
\setlength{\itemsep}{3pt}
\setlength{\parsep}{0pt}
\setlength{\parskip}{0pt}
    \item $Q_{b} = Q \times Q_{ce} \times Q_{mc} \times Q_{ec} \times Q_{sc} \times Q_{i} \times Q_{m} \times Q_{e} \times Q_{s}$
    \item $\Sigma_{b} = (\Sigma - \Sigma_{o,M}) \cup \Sigma_{o,M}^{on} \cup \Sigma_{o,M}^{off} \cup \Delta_{m}^{on} \cup \Delta_{m}^{off} \cup \Sigma_{s,E}^{\#} \cup \Gamma \cup \{stop, decode\}$ 
    \item $\xi_{b}: Q_{b} \times \Sigma_{b} \rightarrow Q_{b}$
    \item $q_{b}^{init} = (q^{init}, q_{ce}^{init}, q_{mc}^{init}, q_{ec}^{init}, q_{sc}^{init}, q_{i}^{init}, q_{m}^{init}, q_{e}^{init}, q_{s}^{init})$
    \item $Q_{b,m} = Q_{G,m} \times Q_{ce,m} \times Q_{mc,m} \times Q_{ec,m} \times Q_{sc} \times Q_{i,m} \times Q_{m,m} \times Q_{e,m} \times Q_{s,m}$ 
\end{itemize}
Based on the closed-loop behavior $\mathcal{B}$, we shall introduce two definitions regarding the opacity and the covertness. Recall that the dynamic mask-control constraint is $\mathcal{C}_{m} = (\Sigma_{m,c}, \Sigma_{m,o})$, the edit function-control constraint is $\mathcal{C}_{e} = (\Sigma_{e,c}, \Sigma_{e,o})$, and the supervisor-control constraint is $\mathcal{C}_{s} = (\Sigma_{s,c}, \Sigma_{s,o})$.

\textbf{Definition IV.1. (Opacity)} Given any plant $G$, command execution component $CE$, dynamic mask constraints $MC$, edit constraints $EC$, supervisor constraints $SC$, and intruder $I$, the combination of the dynamic mask $M$, the edit function $E$, and the supervisor $S$ is an opaque dynamic mask-edit function-supervisor pair w.r.t. $\mathcal{C}_{m}$, $\mathcal{C}_{e}$, and $\mathcal{C}_{s}$ if any state in $Q_{unsafe} = \{(q, q_{ce}, q_{mc}, q_{ec}, q_{sc}, q_{i}, q_{m}, q_{e}, q_{s}) \in Q_{b}|\, q_{i} = q^{unsafe}\}$ is not reachable in $\mathcal{B}$.

\textbf{Definition IV.2. (Covertness)} Given any plant $G$, command execution component $CE$, dynamic mask constraints $MC$, edit constraints $EC$, supervisor constraints $SC$, and intruder $I$, the combination of the dynamic mask $M$, the edit function $E$, and the supervisor $S$ is a covert dynamic mask-edit function-supervisor pair w.r.t. $\mathcal{C}_{m}$, $\mathcal{C}_{e}$, and $\mathcal{C}_{s}$ if any state in $Q_{bad} = \{(q, q_{ce}, q_{mc}, q_{ec}, q_{sc}, q_{i}, q_{m}, q_{e}, q_{s}) \in Q_{b}|\, q_{i} = \varnothing\}$
is not reachable in $\mathcal{B}$.

Next, we shall explain our approach in modeling the problem of co-synthesis of dynamic mask, edit function, and supervisor as a distributed Ramadge-Wonham supervisory control problem.
Since the closed-loop  system is $\mathcal{B} = G||CE||MC||EC||SC||I||M||E||S$, we can view
\[
\mathcal{P} = G||CE||MC||EC||SC||I = (Q_{\mathcal{P}}, \Sigma_{\mathcal{P}}, \xi_{\mathcal{P}}, q_{\mathcal{P}}^{init}, Q_{\mathcal{P},m})
\]
as the new plant and treat $M$, $E$, and $S$ as the distributed supervisor to be synthesized over the distributed control architecture $\mathcal{A}=(\mathcal{C}_{m}, \mathcal{C}_{e}, \mathcal{C}_{s})$. Our goal is to synthesize $M$, $E$, and $S$ such that
\begin{itemize}
\setlength{\itemsep}{3pt}
\setlength{\parsep}{0pt}
\setlength{\parskip}{0pt}
    \item $\mathcal{B}$ is nonblocking and plant $G$ would never reach any state in $Q_{avoid}$.
    \item The combination of $M$, $E$, and $S$ is an opaque and covert dynamic mask-edit function-supervisor pair w.r.t. $\mathcal{C}_{m}$, $\mathcal{C}_{e}$, and $\mathcal{C}_{s}$.
\end{itemize}
Next, we shall briefly review the previous works on distributed supervisor synthesis, which is known to be an undecidable problem \cite{Lin2016Distributed}-\cite{Thistle2005Distributed}. In \cite{Komenda2014CoordinationControl}, the authors propose an approach to synthesize  distributed supervisor by adopting a coordinator. However, in the architecture shown in Fig. \ref{fig:Supervisory control architecture under edit function}, such a coordinator does not exist and this approach cannot be used for the problem to be solved in this work. In \cite{wonham2015supervisory}, the supervisor localization algorithm is proposed, where the observable alphabet for the local supervisor needs to be lifted, which is not applicable for the problem in this work since the observable alphabets of the dynamic mask, edit function, and supervisor are fixed. In \cite{Su2010Aggregative}, an aggregative synthesis method is designed for the distributed supervisory control problem, which always seeks to synthesize a nonblocking local supervisor at each step. Nevertheless, for the dynamic mask, edit function, and supervisor to be synthesized in this work, %the controllable events of any one are uncontrollable for the other two, implying that
it is very likely to generate an empty solution at the first synthesis step by using the algorithm in \cite{Su2010Aggregative}, due to the special distributed control architecture. %{\color{red} this may require more details..as current distributed supervisor synthesis algorithm also deals with this issue. also need to explain the difference with previous paper submitted to TAC. }
%{\color{blue} In \cite{tai2021privacy}, the authors develop two incremental synthesis heuristics, which either firstly synthesize the edit function or the supervisor, to solve the co-synthesis of the edit function and supervisor for opacity enforcement and requirement satisfaction. However, these two heuristics are not applicable to the distributed problem that need to be solved in this work, which are caused by the following two aspects: 1) The incremental synthesis starting from the supervisor cannot work because the additional observation on control commands of the intruder would limit the feasible sensor alteration choices, thus, it is highly possible to generate an empty solution for the dynamic mask and edit function based on the the supervisor synthesized first, e.g., if $S$ always sends a control command only containing one enabled event, then there are no chances for $M$ and $E$ to prevent the intruder from inferring the system secret meanwhile maintaining the covertness against a sensor-actuator intruder, 2) The incremental synthesis from the edit function cannot work because
%}

%{\color{red} we can keep this one not shown here. and only show it in response letter if enquired.. to save space now.. }

Thus, in this work, we shall propose an incremental heuristic to synthesize the dynamic mask, edit function, and supervisor gradually by taking the special structure of the information flow into consideration. The details of the proposed algorithm would be presented in Section \ref{subsec:Incremental Heuristic Synthesis}.

\subsection{Incremental Heuristic Synthesis}
\label{subsec:Incremental Heuristic Synthesis}
We shall briefly explain the main idea of our method. Firstly, we shall synthesize the ensemble of the dynamic mask and the edit function, denoted as $ME$, over the dynamic mask-edit function-control constraint $\mathcal{C}_{me} = (\Sigma_{m,c} \cup \Sigma_{e,c}, \Sigma_{m,o} \cup \Sigma_{e,o})$. We require $ME$ to ensure the marker-reachability, and the opacity and the covertness properties against a  sensor eavesdropping, but command non-eavesdropping  intruder, considering the fact that the sending of control commands are uncontrollable to either the dynamic mask or the edit function. Next, we need to decompose $ME$ into a dynamic mask and an edit function that preserves the marker-reachability. To solve this problem, we shall adopt the technique developed in \cite{neider2012computing} and develop a reduction from the $ME$ decomposition problem  to the SAT problem. Finally, based on the decomposed results, we shall synthesize the supervisor that could ensure the nonblockingness, and the opacity and the covertness against the sensor-actuator eavesdropping intruder. 
Next, we shall present the detailed procedure of the proposed incremental synthesis heuristic:

\textbf{Procedure 1:}
\begin{enumerate}[1.]
\setlength{\itemsep}{3pt}
\setlength{\parsep}{0pt}
\setlength{\parskip}{0pt}
    \item Generate $P_{\widetilde{\Sigma_{o,I}} - \Gamma}(G_{1}) = (Q_{temp}^{w}, \Sigma_{temp}^{w}, \xi_{temp}^{w}, x_{temp}^{w,init})$.
    \item Generate the sensor eavesdropping intruder $I^{w} = (Q_{i}^{w}, \Sigma_{i}^{w}, \xi_{i}^{w}, x_{i}^{w,init}, Q_{i,m}^{w})$, where
    \begin{itemize}
    \setlength{\itemsep}{3pt}
    \setlength{\parsep}{0pt}
    \setlength{\parskip}{0pt}
        \item $Q_{i} = Q_{temp}^{w} \cup \{q^{w,unsafe}\}$
        \item $\Sigma_{i}^{w} = (\Sigma - \Sigma_{o,M} - \Sigma_{s,E}) \cup (\Sigma_{o,M} - \Sigma_{s,E})^{on} \cup \Sigma_{s,E}^{\#} \cup \{decode\}$
        \item  $(\forall q \in Q_{i}^{w}) \, q \in 2^{Q_{sec}} - \{\varnothing\} \Leftrightarrow \xi_{i}^{w}(q, decode) = q^{w,unsafe}$
        \item $(\forall q, q' \in Q_{i}^{w})(\forall \sigma \in \Sigma_{s,E} - \Sigma_{o,M}) \, \xi_{temp}^{w}(q, \sigma) = q' \Leftrightarrow \xi_{i}^{w}(q, \sigma^{\#}) = q'$
        \item $(\forall q, q' \in Q_{i}^{w})(\forall \sigma \in \Sigma_{o,M} \cap \Sigma_{s,E}) \, \xi_{temp}^{w}(q, \sigma^{on}) = q' \Leftrightarrow \xi_{i}^{w}(q, \sigma^{\#}) = q'$
        \item $(\forall q, q' \in Q_{i}^{w})(\forall \sigma \in (\Sigma - \Sigma_{o,M} - \Sigma_{s,E}) \cup (\Sigma_{o,M} - \Sigma_{s,E})^{on}) \, \xi_{temp}^{w}(q, \sigma) = q' \Leftrightarrow \xi_{i}^{w}(q, \sigma) = q'$
        \item $(\forall q \in \{\varnothing, q^{w,unsafe}\})(\forall \sigma \in \Sigma_{i}^{w} - \{decode\})\, \xi_{i}^{w}(q,\sigma)\\ = q$
        \item $q_{i}^{w,init} = q_{temp}^{w,init}$
        \item $Q_{i,m}^{w} = Q_{temp}^{w} - \{\varnothing\}$
    \end{itemize}
    \item Compute $\mathcal{P}_{ME} = G||CE||MC||EC||SC||I^{w} = (Q_{ME}, \Sigma_{ME}, \xi_{ME}, q_{ME}^{init}, Q_{ME,m})$
    \item Generate $\mathcal{P}_{ME}^{r} = (Q_{ME}^{r}, \Sigma_{ME}^{r}, \xi_{ME}^{r}, q_{ME}^{r,init}, Q_{ME,m}^{r})$
    \begin{itemize}
    \setlength{\itemsep}{3pt}
    \setlength{\parsep}{0pt}
    \setlength{\parskip}{0pt}
        \item $Q_{ME}^{r} := Q_{ME} - Q_{1}$, where
        \begin{itemize}
        \setlength{\itemsep}{3pt}
        \setlength{\parsep}{0pt}
        \setlength{\parskip}{0pt}
            \item $Q_{1} := \{(q, q_{ce}, q_{mc}, q_{ec}, q_{sc}, q_{i}) \in Q_{ME}|\, q_{i} = q^{w,unsafe} \vee q_{i} = \varnothing\}$
        \end{itemize}
        \item $\Sigma_{ME}^{r} := \Sigma_{ME}$
        \item $(\forall q, q' \in Q_{ME}^{r})(\forall \sigma \in \Sigma_{ME}^{r})\, \xi_{ME}(q, \sigma) = q' \Leftrightarrow \xi_{ME}^{r}(q, \sigma) = q'$
        \item $q_{ME}^{r,init} := q_{ME}^{init}$
        \item $Q_{ME,m}^{r} := Q_{ME,m} - Q_{1}$
    \end{itemize}
    \item Synthesize a supervisor $ME = (Q_{me}, \Sigma_{me}, \xi_{me}, q_{me}^{init},\\ Q_{me,m})$ over the dynamic mask-edit function-control constraint $\mathcal{C}_{me}$ by treating $\mathcal{P}_{ME}$ as the plant and $\mathcal{P}_{ME}^{r}$ as the requirement such that $\mathcal{P}_{ME}||ME$ is marker-reachable and safe w.r.t. $\mathcal{P}_{ME}^{r}$. If $ME$ exists, go to Step 6; otherwise, end the procedure.
    \item Solve the problem of decomposing\footnote{The details of the decomposition procedure will be explained shortly.} $ME$ into a $k$-bounded dynamic mask $M = (Q_{m}, \Sigma_{m}, \xi_{m}, q_{m}^{init}, Q_{m,m})$ over $\mathcal{C}_{m}$ and an $l$-bounded edit function $E = (Q_{e}, \Sigma_{e}, \xi_{e}, q_{e}^{init}, Q_{e,m})$ over $\mathcal{C}_{e}$ by reducing it to the validity of the formula $\phi_{decompose}^{\bar{M}, \bar{E}, ME, \mathcal{C}_{m}, \mathcal{C}_{e}, k, l}$. If such $M$ and $E$ exist, go to Step 7; otherwise, set $k:= k+1$ and $l:= l+1$ and repeat Step 6.
    \item Compute $\mathcal{P}_{S} = G||CE||MC||EC||SC||I||M||E = (Q_{S}, \Sigma_{S}, \xi_{S}, q_{S}^{init}, Q_{S,m})$
    \item Generate $\mathcal{P}_{S}^{r} = (Q_{S}^{r}, \Sigma_{S}^{r}, \xi_{S}^{r}, q_{S}^{r,init}, Q_{S,m}^{r})$
    \begin{itemize}
    \setlength{\itemsep}{3pt}
    \setlength{\parsep}{0pt}
    \setlength{\parskip}{0pt}
        \item $Q_{S}^{r} := Q_{S} - Q_{2}$
        \begin{itemize}
        \setlength{\itemsep}{3pt}
        \setlength{\parsep}{0pt}
        \setlength{\parskip}{0pt}
            \item $Q_{2} := \{(q, q_{ce}, q_{mc}, q_{ec}, q_{sc}, q_{i}, q_{m}, q_{e}) \in Q_{S}|\, q \\ \in Q_{avoid} \vee q_{i} = q^{unsafe} \vee q_{i} = \varnothing\}$
        \end{itemize}
        \item $\Sigma_{S}^{r} := \Sigma_{S}$
        \item $(\forall q, q' \in Q_{S}^{r})(\forall \sigma \in \Sigma_{S}^{r})\, \xi_{S}(q, \sigma) = q' \Leftrightarrow \xi_{S}^{r}(q, \sigma) = q'$
        \item $q_{S}^{r,init} := q_{S}^{init}$
        \item $Q_{S,m}^{r} := Q_{S,m} - Q_{2}$
    \end{itemize}
    \item Synthesize a supervisor $S = (Q_{s}, \Sigma_{s}, \xi_{s}, q_{s}^{init},Q_{s,m})$ over the supervisor-control constraint $\mathcal{C}_{s} = (\Sigma_{s,c}, \Sigma_{s,o})$ by treating $\mathcal{P}_{S}$ as the plant and $\mathcal{P}_{S}^{r}$ as the requirement such that $\mathcal{P}_{S}||S$ is nonblocking and safe w.r.t. $\mathcal{P}_{S}^{r}$. 
\end{enumerate} 
%Generally speaking, Steps 1-5 focus on the synthesis of the ensemble $ME$, Step 6 decomposes the synthesized $ME$ into the dynamic mask $M$ and edit function $E$, and Steps 6-8 synthesize the supervisor $S$ based on the decomposed $M$ and $E$ at Step 6. 
We shall briefly explain the above procedure. Steps 1-2 generate a weak intruder $I^{w}$ that can only eavesdrop the sensor information. The procedure of constructing $I^{w}$ is similar to that of $I$ in Section \ref{subsec:Intruder}. The only difference is that the state estimator for this weak intruder only requires the structure of $G$, but not the structure $G||CE$ as the sending of control commands are not observable. Based on $I^{w}$, Step 3 and Step 4 compute $\mathcal{P}_{ME}$ and $\mathcal{P}_{ME}^{r}$, respectively, where $\mathcal{P}_{ME}^{r}$ is generated from $\mathcal{P}_{ME}$ by removing the states where $I^{w}$ reaches the state $q^{w,unsafe}$ or $\varnothing$, implying that $I^{w}$ either infers the system secret or discovers the existence of the dynamic mask or the edit function. Then, at Step 5, we compute $ME$ that can ensure the safety w.r.t. $\mathcal{P}_{ME}^{r}$ and the marker-reachability. 
At Step 6, the synthesized $ME$ is decomposed into a dynamic mask $M$ and an edit function $E$, which preserves the marker-reachability, by reducing it to a SAT problem. The details will be introduced later.
In the part regarding the synthesis of a  supervisor, $\mathcal{P}_{S}$ and $\mathcal{P}_{S}^{r}$ are treated as the plant and the requirement, respectively. $\mathcal{P}_{S}$ is generated at Step 7 based on the decomposed $M$ and $E$. At Step 8, the requirement $\mathcal{P}_{S}^{r}$ is generated from $\mathcal{P}_{S}$ by removing the set of states $Q_{2}$, which denotes the states that are not allowed by the user requirement and that the sensor-actuator eavesdropping intruder $I$ has inferred the system secret or discovered the existence of the dynamic mask or edit function. Finally, we compute the supervisor $S$ that can ensure the safety w.r.t. $\mathcal{P}_{S}^{r}$ and the nonblockingness at Step 9. 

%{\color{red} may think about the problem why we do not search for $S$, $E$ and $M$ all by using sat solver instead? 1 reason is sat solver is often more expensive, it is hard to be used for large $S$. Thus, we focus on supervisor decomposition problem for $ME$...something like this. It is possible to do so, but to scale up the size of the instances that can be solved..we adopt this strategy. may later compare both approaches in terms of the synthesis time..}

%{\color{red} how to ensure the resulting $S$, $E$, $M$ together are nonblocking as $M$ and $E$ are obtained from $ME$ without considering $S$. If in the sat constraints, we enforce $S$, $E$ and $M$ are nonblocking, is this going to be expensive?}

%{\color{red} also may explain why not perform incremental synthesis of $S$, $M$, $E$ and why it may be difficult to find a non-empty supervisor based on current synthesis approaches, including the incremental synthesis approach developed recently in the previous paper.}

Next, we shall explain how to formally reduce the decomposition problem of $ME$ to a SAT problem.
For any given parameters $k$ (bounding the size of the dynamic mask) and $l$ (bounding the size of the edit function), the synthesized dynamic mask-edit function $ME$, dynamic mask-control constraint $\mathcal{C}_{m}$, and edit function-control constraint $\mathcal{C}_{e}$, we shall produce a SAT formula $\phi_{decompose}^{\bar{M}, \bar{E}, ME, \mathcal{C}_{m}, \mathcal{C}_{e}, k, l}$ such that $\phi_{decompose}^{\bar{M}, \bar{E}, ME, \mathcal{C}_{m}, \mathcal{C}_{e}, k, l}$ is satisfiable if and only if there exists a $k$-bounded dynamic mask $M$ over $\mathcal{C}_{m}$ and an $l$-bounded edit function $E$ over $\mathcal{C}_{e}$ such that 1) $L(M||E) \subseteq L(ME)$, 2) $L_{m}(M||E) \subseteq L_{m}(ME)$, and 3) $G||CE||MC||EC||SC||I||M||E$ is marker-reachable. Moreover, we can extract a model, if the formula is indeed satisfiable, which can be used to construct a $k$-bounded dynamic mask $M$ and an $l$-bounded edit function $E$ with the prescribed properties.

To ensure $L(M||E) \subseteq L(ME)$ and $L_{m}(M||E) \subseteq L_{m}(ME)$, we need to transform the language inclusion enforcement to the enforcement of the non-reachability of certain states in the synchronous product $M \lVert E \lVert ME$. 
However, in general, $M$, $E$, and $ME$ are partial finite state automata that are not complete. We here remark that enforcing language inclusion is equivalent to enforcing the non-reachability of certain states in the synchronous product, if and only if complete finite state automata are used \cite{lin2019towards}. This trouble can be easily resolved by using the completion $\overline{P}$ of a (partial) finite state automaton $P$. Formally, the completion of any (partial) finite state automaton $P=(V, \Sigma, \xi, v_0, V_m)$ is a complete finite state automaton $\overline{P}=(V \cup \{v_d\}, \Sigma, \overline{\xi}, v_0, V_m)$, where the distinguished state $v_d \notin V$ denotes the added dump state and 
\[
\overline{\xi} = \xi \cup (\{v_d\} \times \Sigma \times \{v_d\}) \cup \bigg \{(v, \sigma, v_d) \bigg|
\begin{aligned}
& \, \neg\xi(v, \sigma)! \wedge v \in V \\ & \wedge \sigma \in \Sigma \end{aligned}
\bigg \}
\]
denotes the transition function. We remark that it is straightforward to recover $P$ from $\overline{P}$; we only need to remove the dump state $v_d$ and the corresponding transitions.  
Thus, to ensure $L(M||E) \subseteq L(ME)$, we only need to ensure in $\overline{M} || \overline{E} || \overline{ME}$ the non-reachability of states for which $\overline{M}$ and $\overline{E}$ are not in the dump states while $\overline{ME}$ is in the dump state. To ensure $L_{m}(M||E) \subseteq L_{m}(ME)$, we only need to ensure in $\overline{M} || \overline{E} || \overline{ME}$ the  non-reachability of states for which $\overline{M}$ and $\overline{E}$ are in some marked states while $\overline{ME}$ is not \cite{neider2012computing}. 

Next, we first explain how the dynamic mask $M$ and the edit function $E$ can be propositionally encoded. Let $M = (Q_{m}, \Sigma_{m}, \xi_{m}, q_{m}^{init}, Q_{m,m})$ denote a $k$-bounded dynamic mask over $\mathcal{C}_{m} = (\Sigma_{m,c}, \Sigma_{m,o})$, where $Q_{m} := \{q_{m}^{0}, q_{m}^{1}, \ldots, q_{m}^{k-1}\}$ consists of $k$ states, $q_{m}^{init} = q_{m}^{0} \in Q_{m}$ is the initial state; the partial transition function $\xi_{m}: Q_{m} \times \Sigma_{m} \rightarrow Q_{m}$ and $Q_{m,m}$ are the parameters that need to be determined to ensure that $M$ is part of a solution of the given instance, if it exists. Our goal is to determine $\overline{M}$, which can be used to recover $M$. We know that $\overline{M}$ is given by the 5-tuple 
\[
(\{q_{m}^{0}, q_{m}^{1}, \ldots, q_{m}^{k-1}, q_{m}^{dump}\}, \Sigma_{m}, \overline{\xi}_{m}, q_{m}^{0}, Q_{m,m})
\]
where $q_{m}^{dump}$ is the added dump state. For convenience, we let $q_{m}^{k} = q_{m}^{dump}$. We then introduce Boolean variables $t_{q_{m}^{i}, \sigma, q_{m}^{j}}^{\overline{M}}$, where $i, j \in [0, k]$ and $\sigma \in \Sigma_{m}$, for the encoding of $\overline{\xi}_{m}$ with the interpretation that $t_{q_{m}^{i}, \sigma, q_{m}^{j}}^{\overline{M}}$ is true if and only if $\overline{\xi}_{m}(q_{m}^{i}, \sigma) = q_{m}^{j}$, and the Boolean variables $m_{q_{m}^{i}}^{\overline{M}}$, where $i \in [0, k-1]$, with the interpretation that $m_{q_{m}^{i}}^{\overline{M}}$ is true if and only if $q_{m}^{i} \in Q_{m,m}$.

Similarly, let $E = (Q_{e}, \Sigma_{e}, \xi_{e}, q_{e}^{init}, Q_{e,m})$ denote an $l$-bounded edit function over $\mathcal{C}_{e} = (\Sigma_{e,c}, \Sigma_{e,o})$, where $Q_{e} := \{q_{e}^{0}, q_{e}^{1}, \ldots, q_{e}^{l-1}\}$ consists of $l$ states, $q_{e}^{init} = q_{e}^{0} \in Q_{e}$ is the initial state; the partial transition function $\xi_{e}: Q_{e} \times \Sigma_{e} \rightarrow Q_{e}$ and $Q_{e,m}$ are the parameters that need to be determined. We know that $\overline{E}$ is given by the 5-tuple 
\[
(\{q_{e}^{0}, q_{e}^{1}, \ldots, q_{e}^{l-1}, q_{e}^{dump}\}, \Sigma_{e}, \overline{\xi}_{e}, q_{e}^{0}, Q_{e,m})
\]
where $q_{e}^{dump}$ is the added dump state. For convenience, we let $q_{e}^{l} = q_{e}^{dump}$. We then introduce Boolean variables $t_{q_{e}^{i}, \sigma, q_{e}^{j}}^{\overline{E}}$, where $i, j \in [0, l]$ and $\sigma \in \Sigma_{e}$, for the encoding of $\overline{\xi}_{e}$ with the interpretation that $t_{q_{e}^{i}, \sigma, q_{e}^{j}}^{\overline{E}}$ is true if and only if $\overline{\xi}_{e}(q_{e}^{i}, \sigma) = q_{e}^{j}$, and the Boolean variables $m_{q_{e}^{i}}^{\overline{E}}$, where $i \in [0, l-1]$, with the interpretation that $m_{q_{e}^{i}}^{\overline{E}}$ is true if and only if $q_{e}^{i} \in Q_{e,m}$.

We encode the fact that $\overline{\xi}_{m}$ and $\overline{\xi}_{e}$ are transition functions using the following constraints:
\begin{enumerate}
\setlength{\itemsep}{3pt}
\setlength{\parsep}{0pt}
\setlength{\parskip}{0pt}
\item [1.] $t_{q_{m}^{k}, \sigma, q_{m}^{k}}^{\overline{M}}$, for each $\sigma \in \Sigma_{m}$
\item [2.] $t_{q_{e}^{l}, \sigma, q_{e}^{l}}^{\overline{E}}$, for each $\sigma \in \Sigma_{e}$
\item [3.] $\neg t_{q_{m}^{i}, \sigma, q_{m}^{j}}^{\overline{M}} \vee \neg t_{q_{m}^{i}, \sigma, q_{m}^{h}}^{\overline{M}}$, for each $i \in [0, k-1]$, each $\sigma \in \Sigma_{m}$ and each $j \neq h \in [0, k]$
\item [4.] $\neg t_{q_{e}^{i}, \sigma, q_{e}^{j}}^{\overline{E}} \vee \neg t_{q_{e}^{i}, \sigma, q_{e}^{h}}^{\overline{E}}$, for each $i \in [0, l-1]$, each $\sigma \in \Sigma_{e}$ and each $j \neq h \in [0, l]$
\item [5.] $\bigvee_{j \in [0, k]}t_{q_{m}^{i}, \sigma, q_{m}^{j}}^{\overline{M}}$, for each $i \in [0, k-1]$ and each $\sigma \in \Sigma_{m}$ 
\item [6.] $\bigvee_{j \in [0, l]}t_{q_{e}^{i}, \sigma, q_{e}^{j}}^{\overline{E}}$, for each $i \in [0, l-1]$ and each $\sigma \in \Sigma_{e}$ 
\end{enumerate}
We here remark that Constraints 1 and 2 encode the fact that any event leads to a self-loop at the dump state $q_{m}^{k}$ or $q_{m}^{l}$. Constraints 3 and 4 are then imposed to ensure that $\overline{\xi}_{m}$ and $\overline{\xi}_{e}$ are deterministic. Constraints 5 and 6 are imposed to ensure that $\overline{\xi}_{m}$ and $\overline{\xi}_{e}$ are total. 
Together, they will ensure that $\overline{\xi}_{m}$ and $\overline{\xi}_{e}$ are transition functions and thus $\overline{M}$ and $\overline{E}$ are complete finite state automata. Let $\phi_{k,l}^{\overline{M},\overline{E}, fsa}$ denote the resultant formula obtained after combining the above Constraints 1-6 conjunctively.

With the above constraints, we can encode the fact that $M$ is a finite state dynamic mask over $\mathcal{C}_{m} = (\Sigma_{m,c}, \Sigma_{m,o})$ and $E$ is a finite state edit function over $\mathcal{C}_{e} = (\Sigma_{e,c}, \Sigma_{e,o})$ using the following constraints. 
\begin{enumerate}
\setlength{\itemsep}{3pt}
\setlength{\parsep}{0pt}
\setlength{\parskip}{0pt}
\item [7.] $\bigvee_{j \in [0, k-1]} t_{q_{m}^{i}, \sigma, q_{m}^{j}}^{\overline{M}}$ for each $i \in [0, k-1]$ and each $\sigma \in \Sigma_{m,uc}$
\item [8.] $(\bigvee_{j \in [0, k-1]} t_{q_{m}^{i}, \sigma, q_{m}^{j}}^{\overline{M}}) \Rightarrow t_{q_{m}^{i}, \sigma, q_{m}^{i}}^{\overline{M}}$ for each $i \in [0, k-1]$ and each $\sigma \in \Sigma_{m,uo}$
\item [9.] $\bigvee_{j \in [0, l-1]} t_{q_{e}^{i}, \sigma, q_{e}^{j}}^{\overline{E}}$ for each $i \in [0, l-1]$ and each $\sigma \in \Sigma_{e,uc}$
\item [10.] $(\bigvee_{j \in [0, l-1]} t_{q_{e}^{i}, \sigma, q_{e}^{j}}^{\overline{E}}) \Rightarrow t_{q_{e}^{i}, \sigma, q_{e}^{i}}^{\overline{E}}$ for each $i \in [0, l-1]$ and each $\sigma \in \Sigma_{e,uo}$
\end{enumerate}
Constraints 7 and 8 are imposed to ensure the M-controllability and M-observability. Constraints 9 and 10 are imposed to ensure the E-controllability and E-observability. Let $\phi_{k,l}^{\overline{M},\overline{E}, con\_obs}$ denote the resultant formula after combining Constraints 7-10 conjunctively.

As we have discussed before, to encode $L(M||E) \subseteq L(ME)$ and $L_{m}(M||E) \subseteq L_{m}(ME)$, we need to first obtain the completion $\overline{ME} = (Q_{me} \cup \{q_{me}^{dump}\}, \Sigma_{me}, \overline{\xi}_{me}, q_{me}^{init}, Q_{me,m})$ of $ME$, with the added dump state $q_{me}^{dump}$. Then we need to track the synchronous product $\overline{M} || \overline{E} || \overline{ME}$ to ensure the non-reachability of states for which 1) $\overline{M}$ and $\overline{E}$ are not in the dump states while $\overline{ME}$ is in the dump state, and 2) $\overline{M}$ and $\overline{E}$ are in some marked states while $\overline{ME}$ is not. Thus, we only need to ensure the existence of an inductive invariant $\mathcal{I} \subseteq (Q_{m} \cup \{q_{m}^{k}\}) \times (Q_{e} \cup \{q_{e}^{l}\}) \times (Q_{me} \cup \{q_{me}^{dump}\})$ such that
\begin{enumerate}
\setlength{\itemsep}{3pt}
\setlength{\parsep}{0pt}
\setlength{\parskip}{0pt}
    \item [a)] $(q_{m}^{0}, q_{e}^{0}, q_{me}^{init}) \in \mathcal{I}$,
    \item [b)] for any $ (q_{m}, q_{e}, q_{me}) \in \mathcal{I}$ and any $\sigma \in \Sigma_{m} = \Sigma_{e} = \Sigma_{me}$, $\overline{\xi}_{m} || \overline{\xi}_{e} || \overline{\xi}_{me}((q_{m}, q_{e}, q_{me}), \sigma) \in \mathcal{I}$, 
    \item [c)] $\{(q_{m}, q_{e}, q_{me}) \in Q_{m} \times Q_{e} \times \{q_{me}^{dump}\}\} \cap \mathcal{I} = \varnothing$.
    \item [d)] $\{(q_{m}, q_{e}, q_{me}) \in Q_{m,m} \times Q_{e,m} \times (Q_{me} - Q_{me,m})\} \cap \mathcal{I} = \varnothing$.
\end{enumerate}
Rule a) and Rule b) ensures that $\mathcal{I}$ is an inductive invariant and thus an over-approximation of the set of reachable states; Rule c) ensures the non-reachability of states for which $\overline{M}$ and $\overline{E}$ are not in the dump states while $\overline{ME}$ is in the dump state; Rule d) ensures the non-reachability of states for which $\overline{M}$ and $\overline{E}$ are in some marked states while $\overline{ME}$ is not. Since the state space of $\overline{M} || \overline{E} \lVert \overline{ME}$ is finite, we can propositionally encode $\mathcal{I}$, by using a Boolean variable for each state $(q_{m}, q_{e}, q_{me})$ of $(Q_{m} \cup \{q_{m}^{k}\}) \times (Q_{e} \cup \{q_{e}^{l}\}) \times (Q_{me} \cup \{q_{me}^{dump}\})$ to encode whether $(q_{m}, q_{e}, q_{me}) \in \mathcal{I}$. 
For $\overline{M} || \overline{E} || \overline{ME}$, we now introduce the auxiliary Boolean variables $r_{q_{m}, q_{e}, q_{me}}$, where $q_{m} \in Q_{m} \cup \{q_{m}^{k}\}$, $q_{e} \in Q_{e} \cup \{q_{e}^{l}\}$, and $q_{me} \in Q_{me} \cup \{q_{me}^{dump}\}$, with the interpretation that $r_{q_{m}, q_{e}, q_{me}}$ is true iff $(q_{m}, q_{e}, q_{me}) \in \mathcal{I}$. Then, we have the following constraints:
\begin{enumerate}
\setlength{\itemsep}{3pt}
\setlength{\parsep}{0pt}
\setlength{\parskip}{0pt}
    \item [11.] $r_{q_{m}^{0}, q_{e}^{0}, q_{me}^{init}}$
    \item [12.] $r_{q_{m}^{i}, q_{e}^{j}, q_{me}} \wedge t_{q_{m}^{i}, \sigma, q_{m}^{i'}}^{\overline{M}} \wedge t_{q_{e}^{j}, \sigma, q_{e}^{j'}}^{\overline{E}} \Rightarrow r_{q_{m}^{i'}, q_{e}^{j'}, q_{me}'}$, for each $i, i' \in [0, k]$, each $j, j' \in [0, l]$, each $q_{me}, q_{me}' \in Q_{me} \cup \{q_{me}^{dump}\}$ and each $\sigma \in \Sigma_{m} = \Sigma_{e} = \Sigma_{me}$ such that $\overline{\xi}_{me}(q_{me}, \sigma) = q_{me}'$.
    \item [13.] $\bigwedge_{q_{m} \in Q_{m}, q_{e} \in Q_{e}} \neg r_{q_{m}, q_{e}, q_{me}^{dump}}$
    \item [14.] $m_{q_{m}^{i}}^{\overline{M}} \wedge m_{q_{e}^{j}}^{\overline{E}} \Rightarrow \bigwedge_{q_{me} \in (Q_{me} - Q_{me,m}) \cup \{q_{me}^{dump}\}}\neg r_{q_{m}^{i}, q_{e}^{j}, q_{me}}$ for each $i \in [0,k-1]$ and each $j \in [0,l-1]$.
\end{enumerate}
Intuitively, Constraints 11, 12, 13, and 14 are used to encode Rule a), Rule b), Rule c), and Rule d), respectively. Let $\phi_{k, l}^{\overline{M}, \overline{E}, \overline{ME}}$ denote the resultant formula after combining Constraints 11-14 conjunctively. Then, $\phi_{k, l}^{\overline{M}, \overline{E}, \overline{ME}}$ enforces $L(M||E) \subseteq L(ME)$ and $L_{m}(M||E) \subseteq L_{m}(ME)$.

Next, we show how to enforce $G||CE||MC||EC||SC||I||\\M||E$ is marker-reachable. We adopt a variation of the bounded model checking technique of \cite{biere2003bounded} and denote $G||CE||MC||EC||SC||I = (\widetilde{Q}, \widetilde{\Sigma} = \Sigma_{m} = \Sigma_{e}, \widetilde{\xi}, \widetilde{q^{0}}, \widetilde{Q}_{m})$.
We introduce the auxiliary Boolean variables $r_{q_{m}, q_{e}, \widetilde{q}}^{t}$, where $0 \leq t \leq kl|\widetilde{Q}|-1$, $q_{m} \in Q_{m}$, $q_{e} \in Q_{e}$, and $\widetilde{q} \in \widetilde{Q}$ with the interpretation that $r_{q_{m}, q_{e}, \widetilde{q}}^{t}$ is true iff the state $(q_{m}, q_{e}, \widetilde{q}) \in Q_{m} \times Q_{e} \times \widetilde{Q}$ can be reached from the initial state $(q_{m}^{0}, q_{e}^{0}, \widetilde{q^{0}}) \in Q_{m} \times Q_{e} \times \widetilde{Q}$ within $t$ transition steps in $M||E||G||CE||MC||EC||SC||I$, counting the stuttering steps. We have the following constraints.
\begin{enumerate}
\setlength{\itemsep}{3pt}
\setlength{\parsep}{0pt}
\setlength{\parskip}{0pt}
    \item [15.] $r_{q_{m}^{0}, q_{e}^{0}, \widetilde{q^{0}}}^{0}$
    \item [16.] $\neg r_{q_{m}^{i}, q_{e}^{j}, \widetilde{q}}^{0}$ for each $i \in [0, k-1]$, each $j \in [0, l-1]$, and each $\widetilde{q} \in \widetilde{Q}$ with $(q_{m}^{i}, q_{e}^{j}, \widetilde{q}) \neq (q_{m}^{0}, q_{e}^{0}, \widetilde{q^{0}})$
    \item [17.] $r_{q_{m}^{i'}, q_{e}^{j'}, \widetilde{q'}}^{t+1} \Leftrightarrow \bigvee_{i \in [0, k-1], j \in [0, l-1], \widetilde{q} \in \widetilde{Q}, \sigma \in \Sigma_{m}, \widetilde{\xi}(\widetilde{q}, \sigma) = \widetilde{q'}}\\(r_{q_{m}^{i}, q_{e}^{j}, \widetilde{q}}^{t} \wedge t_{q_{m}^{i}, \sigma, q_{m}^{i'}}^{\overline{M}} \wedge t_{q_{e}^{j}, \sigma, q_{e}^{j'}}^{\overline{E}}) \vee r_{q_{m}^{i'}, q_{e}^{j'}, \widetilde{q'}}^{t}$ for each $i' \in [0, k-1]$, each $j' \in [0, l-1]$, each $\widetilde{q'} \in \widetilde{Q}$, and each $t \in [0, kl|\widetilde{Q}|-2]$
    \item [18.]
     $\bigvee_{i\in [0,k-1], j\in [0,l-1] }(m_{q_{m}^{i}}^{\overline{M}} \wedge m_{q_{e}^{j}}^{\overline{E}} \wedge \bigvee_{t \in [0, kl|\widetilde{Q}|-1], \widetilde{q} \in \widetilde{Q}_{m}} \\ r_{q_{m}^{i}, q_{e}^{j}, \widetilde{q}}^{t}$) 
\end{enumerate}
Intuitively, Constraints 15 and 16 ensure the interpretation of $r_{q_{m}, q_{e}, \widetilde{q}}^{t}$ is correct for the base case $t = 0$. Constraint 17 inductively enforce the correctness of the interpretation of $r_{q_{m}, q_{e}, \widetilde{q}}^{t+1}$, based on the correctness of the interpretation of $r_{q_{m}, q_{e}, \widetilde{q}}^{t}$. Thus, Constraints 15, 16 and 17 together enforce the correctness of the interpretation of $r_{q_{m}, q_{e}, \widetilde{q}}^{t}$, for each $0 \leq t \in kl|\widetilde{Q}| - 1$, $q_{m} \in Q_{m}$, $q_{e} \in Q_{e}$, and $\widetilde{q} \in \widetilde{Q}$. Constraint 18 ensures that some marked state is reachable in $M||E||G||CE||MC||EC||SC||I$.
Let $\phi_{reachable}^{\overline{M}, \overline{E}, k, l}$ denote the resultant formula after combining Constraints 15-18 conjunctively.

Now, we let $X := \{t_{q_{m}^{i}, \sigma, q_{m}^{j}}^{\overline{M}}| i, j \in [0, k] \wedge \sigma \in \Sigma_{m}\} \cup \{m_{q_{m}^{i}}^{\overline{M}}|i \in [0, k-1]\}$ denote the list of Boolean variables that encodes the dynamic mask $M$ and let $Y := \{t_{q_{e}^{i}, \sigma, q_{e}^{j}}^{\overline{E}}| i, j \in [0, l] \wedge \sigma \in \Sigma_{e}\} \cup \{m_{q_{e}^{i}}^{\overline{E}}|i \in [0, l-1]\}$
denote the list of Boolean variables that encodes the edit function $E$. 
Let $Z := \{r_{q_{m}^{i}, q_{e}^{j}, q_{me}}| i \in [0, k] \wedge j \in [0, l] \wedge q_{me} \in Q_{me} \cup \{q_{me}^{dump}\}\}$ denote the auxiliary Boolean variables for $\phi_{k, l}^{\overline{M}, \overline{E}, \overline{ME}}$. 
Let $R^{reachable}:=\{r_{q_{m}, q_{e}, \widetilde{q}}^{t} | q_{m} \in Q_{m}, q_{e} \in Q_{e}, \widetilde{q} \in \widetilde{Q}, t \in [0, kl|\widetilde{Q}|-1]\}$ denote the auxiliary Boolean variables for formula $\phi_{reachable}^{\overline{M}, \overline{E}, k, l}$
Then, the problem of decomposing $ME$ into a $k$-bounded dynamic mask $M$ and an $l$-bounded edit function $E$ is reduced to the satisfiability of the following SAT formula $\phi_{decompose}^{\bar{M}, \bar{E}, ME, \mathcal{C}_{m}, \mathcal{C}_{e}, k, l} :=$
\[
\begin{aligned}
&\phi_{k,l}^{\overline{M},\overline{E}, fsa} \wedge \phi_{k,l}^{\overline{M},\overline{E}, con\_obs} \wedge \phi_{k, l}^{\overline{M}, \overline{E}, \overline{ME}}  &\wedge  \phi_{reachable}^{\overline{M}, \overline{E}, k, l}
%\exists X, \exists Y, (&\phi_{k,l}^{\overline{M},\overline{E}, fsa} \wedge \phi_{k,l}^{\overline{M},\overline{E}, con\_obs} \wedge (\exists Z, \phi_{k, l}^{\overline{M}, \overline{E}, \overline{ME}}) \\ &\wedge (\exists R^{reachable}, \phi_{reachable}^{\overline{M}, \overline{E}, k, l}))
\end{aligned}
\]

\textbf{Theorem 4.1} Given the synthesized dynamic mask-edit function $ME$, dynamic mask-control constraint $\mathcal{C}_{m}$, and edit function-control constraint $\mathcal{C}_{e}$, there exist a $k$-bounded dynamic mask $M$ over $\mathcal{C}_{m}$ and an $l$-bounded edit function $E$ over $\mathcal{C}_{e}$ such that 1) $L(M||E) \subseteq L(ME)$, 2) $L_{m}(M||E) \subseteq L_{m}(ME)$, and 3) $G||CE||MC||EC||SC||I||M||E||S$ is marker-reachable iff $\phi_{decompose}^{\bar{M}, \bar{E}, ME, \mathcal{C}_{m}, \mathcal{C}_{e}, k, l}$ is satisfiable.

\emph{Proof:} This follows directly from the interpretations of the Boolean variables in $X, Y, Z, R^{reachable}$ and the constructions of the formulas $\phi_{k,l}^{\overline{M},\overline{E}, fsa}$, $\phi_{k,l}^{\overline{M},\overline{E}, con\_obs}$, $\phi_{k, l}^{\overline{M}, \overline{E}, \overline{ME}}$, and $\phi_{reachable}^{\overline{M}, \overline{E}, k, l}$. \hfill $\blacksquare$

\textbf{Theorem 4.2} Given any plant $G$, command execution component $CE$, dynamic mask constraints $MC$, edit constraints $EC$, supervisor constraints $SC$, and intruder $I$, the computed $M$, $E$, and $S$ in \textbf{Procedure 1}, if they exist, satisfy the following goals:
\begin{itemize}
\setlength{\itemsep}{3pt}
\setlength{\parsep}{0pt}
\setlength{\parskip}{0pt}
    \item $G||CE||MC||EC||SC||I||M||E||S$ is nonblocking and any state in $\{(q, q_{ce}, q_{mc}, q_{ec}, q_{sc}, q_{i}, q_{m}, q_{e}, q_{s}) \in Q_{b}|\, q \in Q_{avoid}\}$ is not reachable in $G||CE||MC||EC||\\SC||I||M||E||S$.
    \item The combination of $M$, $E$, and $S$ is an opaque and covert dynamic mask-edit function-supervisor pair w.r.t. the dynamic mask-control constraint $\mathcal{C}_{m}$, edit function-control constraint $\mathcal{C}_{e}$, and supervisor-control constraint $\mathcal{C}_{s}$.
\end{itemize}
\emph{Proof:} Based on Step 9 of \textbf{Procedure 1}, the synthesized $S$ should satisfy that $G||CE||MC||EC||SC||I||M||E||S$ is nonblocking. In addition, at Step 8, the set of states $Q_{2}$ has been removed from $\mathcal{P}_{S}^{r}$, i.e., they are treated as ``bad'' states in the synthesis of $S$. Thus, any state in $\{(q, q_{ce}, q_{mc} q_{ec}, q_{sc}, q_{i}, q_{m}, q_{e}, q_{s}) \in Q_{b}|\, q \in Q_{avoid} \vee q_{i} = q^{unsafe} \vee q_{i} = \varnothing\}$ is not reachable in $G||CE||MC||EC||SC||I||M||E||S$. 
Based on Definition IV.1 and IV.2, the combination of the computed $M$, $E$, and $S$ in \textbf{Procedure 1} is an opaque and covert dynamic mask-edit function-supervisor pair. This completes the proof. \hfill $\blacksquare$

Next, we shall analyze the computational complexity of \textbf{Procedure 1}. Assuming $k$ and $l$ are initially set to large enough values to avoid the iterative increment of $k$ and $l$ in Step 6 of {\bf Procedure 1}, then, by adopting the normality based synthesis approach in \cite{WangStateControl2018}, the complexity of \textbf{Procedure 1} is $O(|\Sigma_{ME}|2^{|Q_{ME}|} + 2^{N} + |\Sigma_{s}||Q_{s}|^{2}4^{|Q_{s}|}) = O(2^{N})$, where
\begin{itemize}
\setlength{\itemsep}{3pt}
\setlength{\parsep}{0pt}
\setlength{\parskip}{0pt}
    \item $|Q_{ME}| = |Q| \times |Q_{ce}| \times |Q_{mc}| \times |Q_{ec}| \times |Q_{sc}| \times |Q_{i}^{w}|$ with  $|Q_{i}^{w}| \leq 2^{|Q|} + 1$.
    \item $N = (k+1)^{2}|\Sigma_{m}| + k + (l+1)^{2}|\Sigma_{e}| + l + (k+1)(l+1)(|Q_{me}|+1) + k^{2}l^{2}|\widetilde{Q}|^{2}$ denotes the number of Boolean variables in $X$, $Y$, $Z$, and $R^{reachable}$. 
    \item $|\widetilde{Q}| = |Q| \times |Q_{ce}| \times |Q_{mc}| \times |Q_{ec}| \times |Q_{sc}| \times |Q_{i}|$
    \item $|Q_{s}| = |Q| \times |Q_{ce}| \times |Q_{mc}| \times |Q_{ec}| \times |Q_{sc}| \times |Q_{i}| \times |Q_{m}| \times |Q_{e}|$
\end{itemize}

%%%%%%%%%%%%%%%%%%%%%%%%%%%%%%%%%%%%%%%%%%%%%%%%%%%%%%%%%%%%%%%%%%%%%%%%%%%%%%%%

\section{Example}
\label{sec:example}
%In this section, we shall present an example to show the effectiveness of the proposed method to synthesize edit function and supervisor for opacity enforcement.

\begin{figure}[htbp]
\begin{center}
\includegraphics[height=5.4cm]{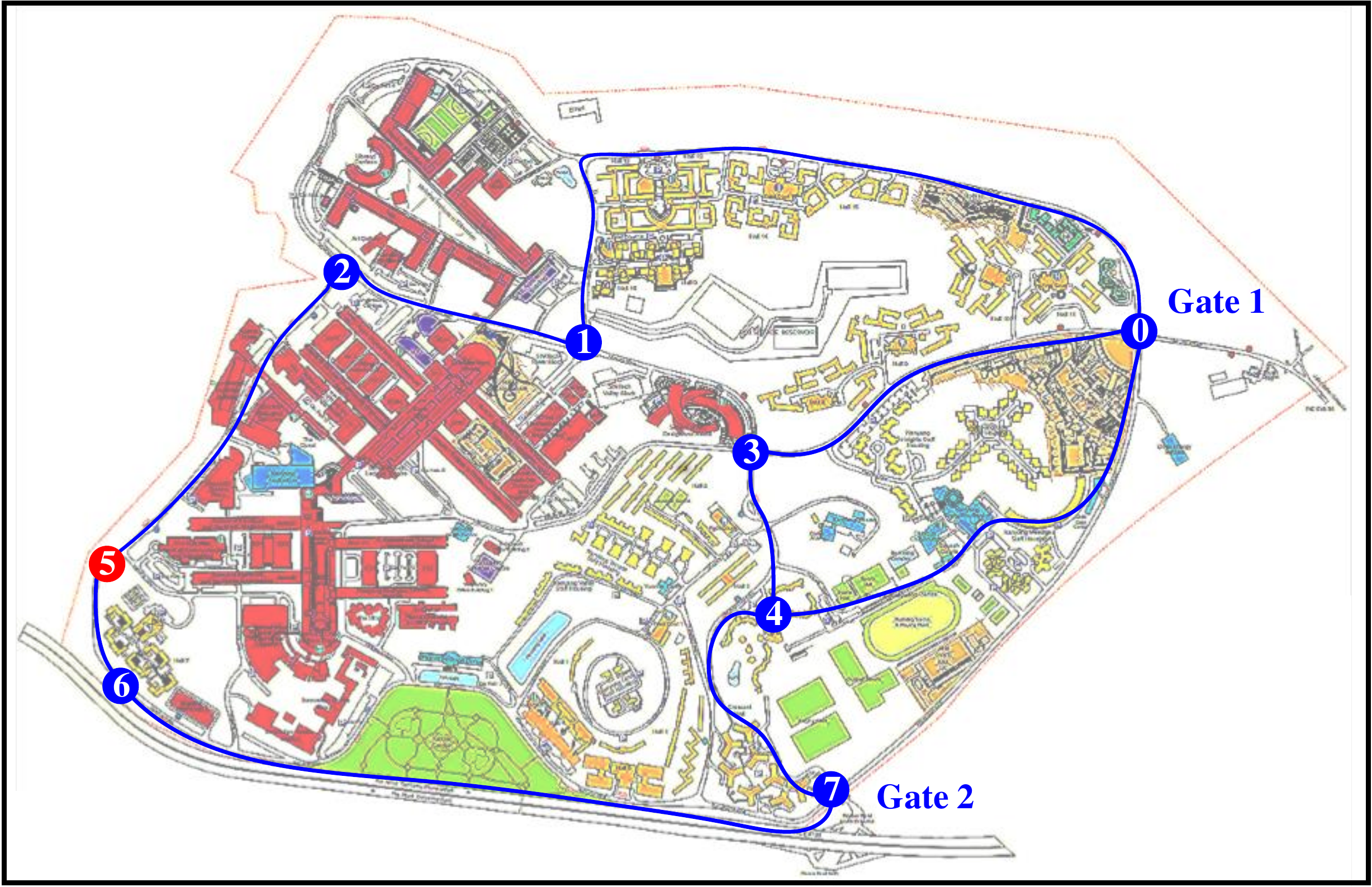}   
\caption{The Nanyang Technological University campus map}
\label{fig:The Nanyang Technological University campus map}
\end{center}        
\end{figure}

\textbf{Example 5.1} We adapt the location-based privacy example of \cite{Wu2014LocationPrivacy} for an illustration. In this example, a vehicle is required to enter the campus from Gate 1, and leave the campus from Gate 2 after completing some transportation task.
The Nanyang Technological University campus map is shown in Fig. \ref{fig:The Nanyang Technological University campus map}, where we discretize the model by selecting eight locations as states, marked by blue circles, and several connection routes between those locations, marked by blue lines. Location (state) 5 represents the EEE building, which is the secret location (state) that the intruder intends to infer. 

In this example, $\Delta = \{s_{1}, s_{2}, s_{3}, s_{4}, s_{5}, s_{6}\}$. $\Sigma = \{a, b, c, d, e, f\}$. $\Sigma_{s}^{1} = \{a\}$. $\Sigma_{s}^{2} = \{b\}$. $\Sigma_{s}^{3} = \{c\}$. $\Sigma_{s}^{4} = \{d\}$. $\Sigma_{s}^{5} = \{e\}$. $\Sigma_{s}^{6} = \{f\}$. $\Sigma_{uc} = \{d, e, f\}$. $\Sigma_{o} = \Sigma$. $\Delta_{m} = \{s_{3}, s_{5}\}$. $\Sigma_{o,M} = \{c, e\}$. $\Sigma_{o,E} = \{a, b, c, d, e, f\}$. $\Sigma_{s,E} = \{b, c\}$. $\Sigma_{o,I} = \{b, c, d, e, f\}$. $U = 1$.
$\Gamma = \{v_{1}, v_{2}, v_{3}, v_{4}, v_{5}, v_{6}, v_{7}\}$, where $v_{1} = \{a\}$, $v_{2} = \{b\}$, $v_{3} = \{c\}$, $v_{4} = \{a, b\}$, $v_{5} = \{a, c\}$, $v_{6} = \{b, c\}$, and $v_{7} = \{a, b, c\}$. $\Gamma_{o} = \{v_{2}, v_{3}, v_{4}, v_{5}, v_{6}, v_{7}\}$. 

\begin{figure}[htbp]
\centering
\subfigure[]{
\begin{minipage}[t]{0.3\linewidth}
\centering
\includegraphics[height=1.1in]{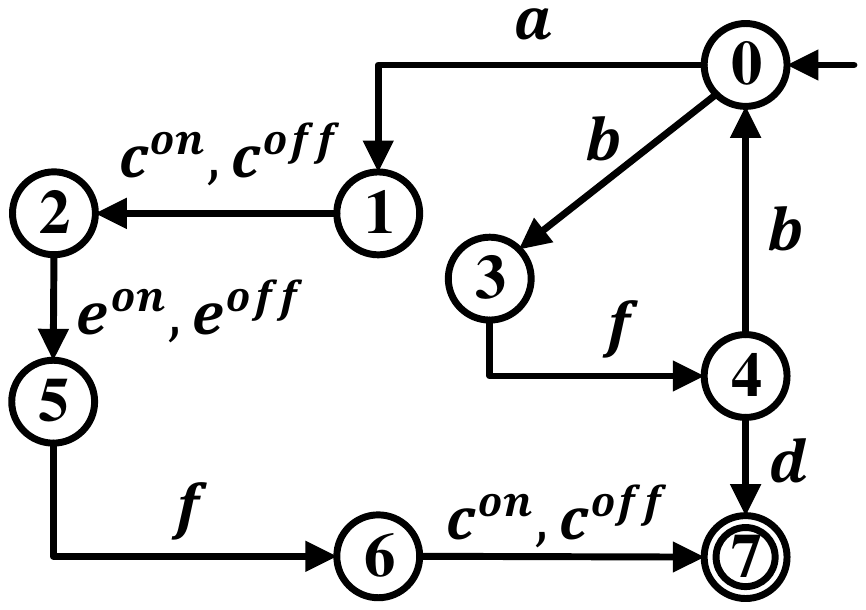}
\end{minipage}
}
\hfill
\subfigure{
\begin{minipage}[t]{0.4\linewidth}
\centering
\includegraphics[height=1.1in]{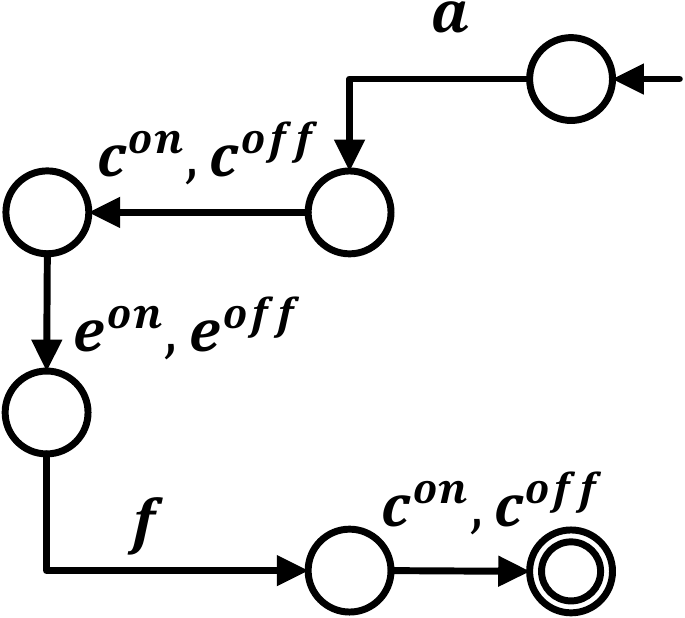}
\end{minipage}
}
\centering
\caption{(a) Plant $G$. (b) Requirement of the plant $G$.}
\label{fig:Plant and Requirement}
\end{figure}

\begin{figure}[htbp]
\begin{center}
\includegraphics[height=6cm]{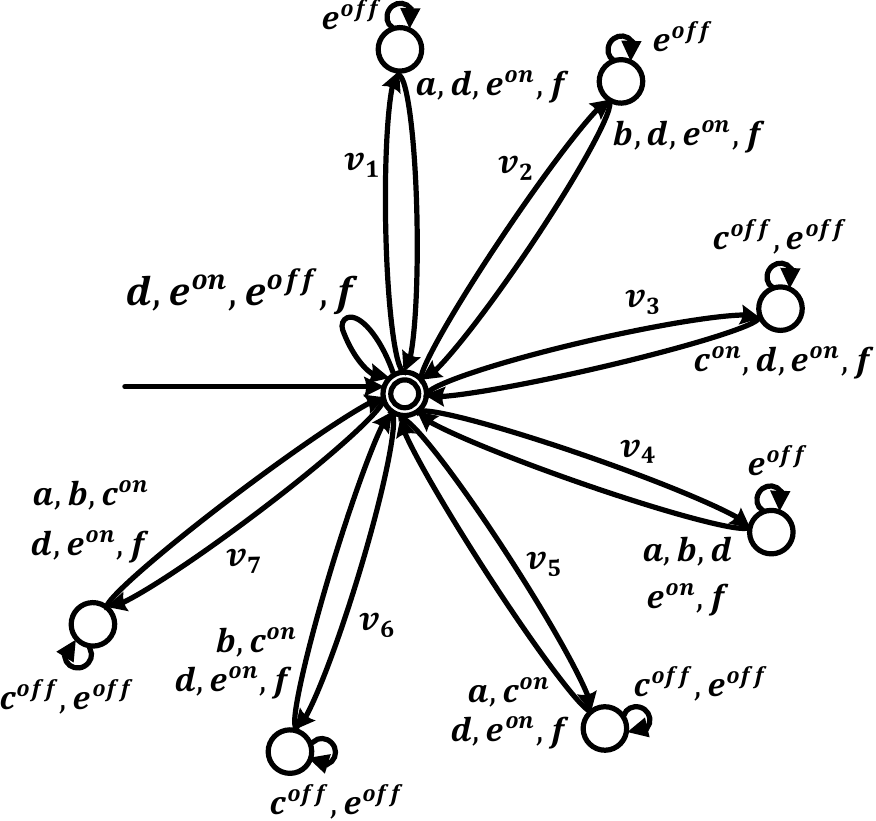}   
\caption{Command execution component $CE$}
\label{fig:Command execution CE}
\end{center}        
\end{figure}

\begin{figure}[htbp]
\begin{center}
\includegraphics[height=2.6cm]{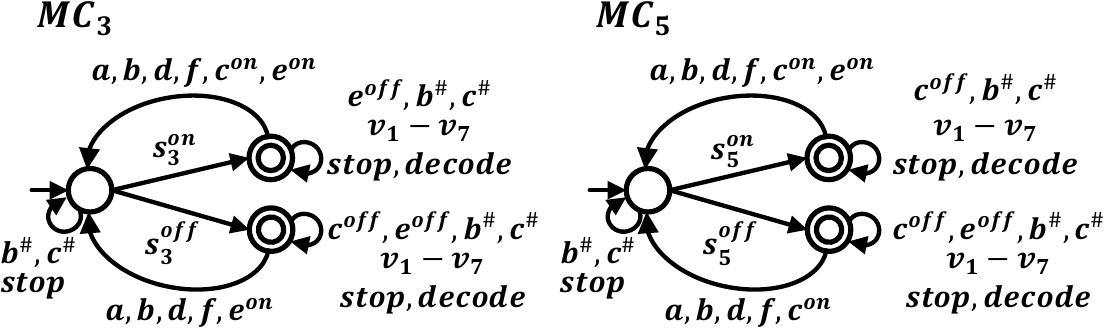}   
\caption{Mask constraints $MC_{3}$ and $MC_{5}$}
\label{fig:Mask constraints MC}
\end{center}        
\end{figure}

\begin{figure}[htbp]
\begin{center}
\includegraphics[height=3cm]{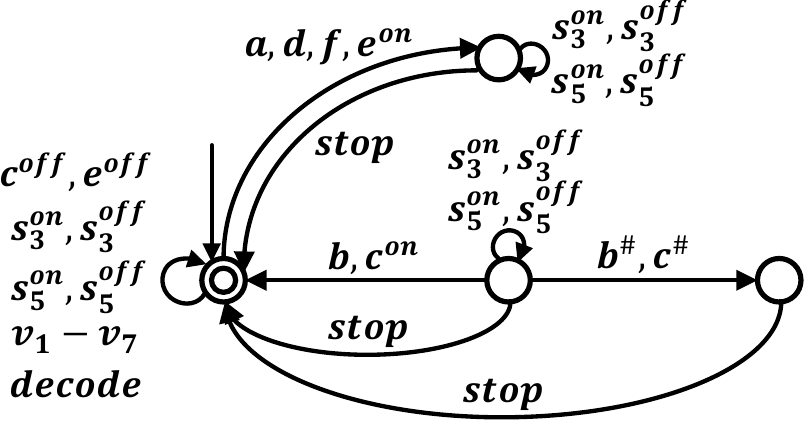}   
\caption{Edit constraints $EC$}
\label{fig:Edit constraints EC}
\end{center}        
\end{figure}

\begin{figure}[htbp]
\begin{center}
\includegraphics[height=3.3cm]{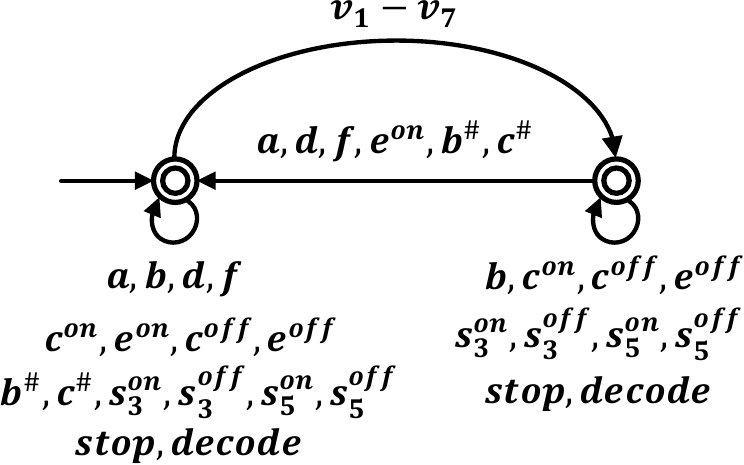}   
\caption{Supervisor constraints $SC$}
\label{fig:Supervisor constraints SC}
\end{center}        
\end{figure}

\begin{figure*}[htbp]
\begin{center}
\includegraphics[height=13.5cm]{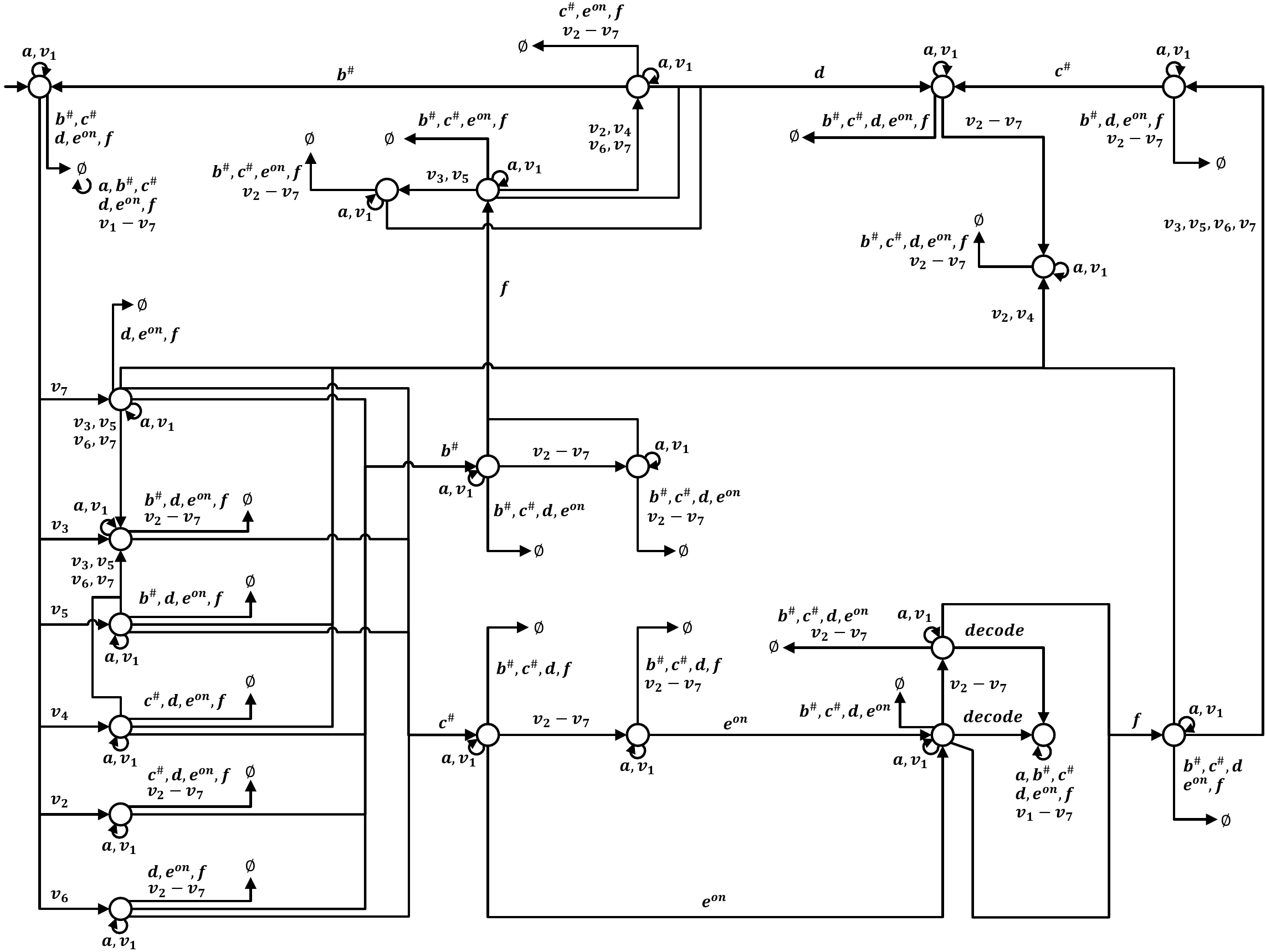}   
\caption{Sensor-actuator eavesdropping intruder $I$}
\label{fig:intruder I}
\end{center}        
\end{figure*}

\begin{figure}[htbp]
\begin{center}
\includegraphics[height=5.2cm]{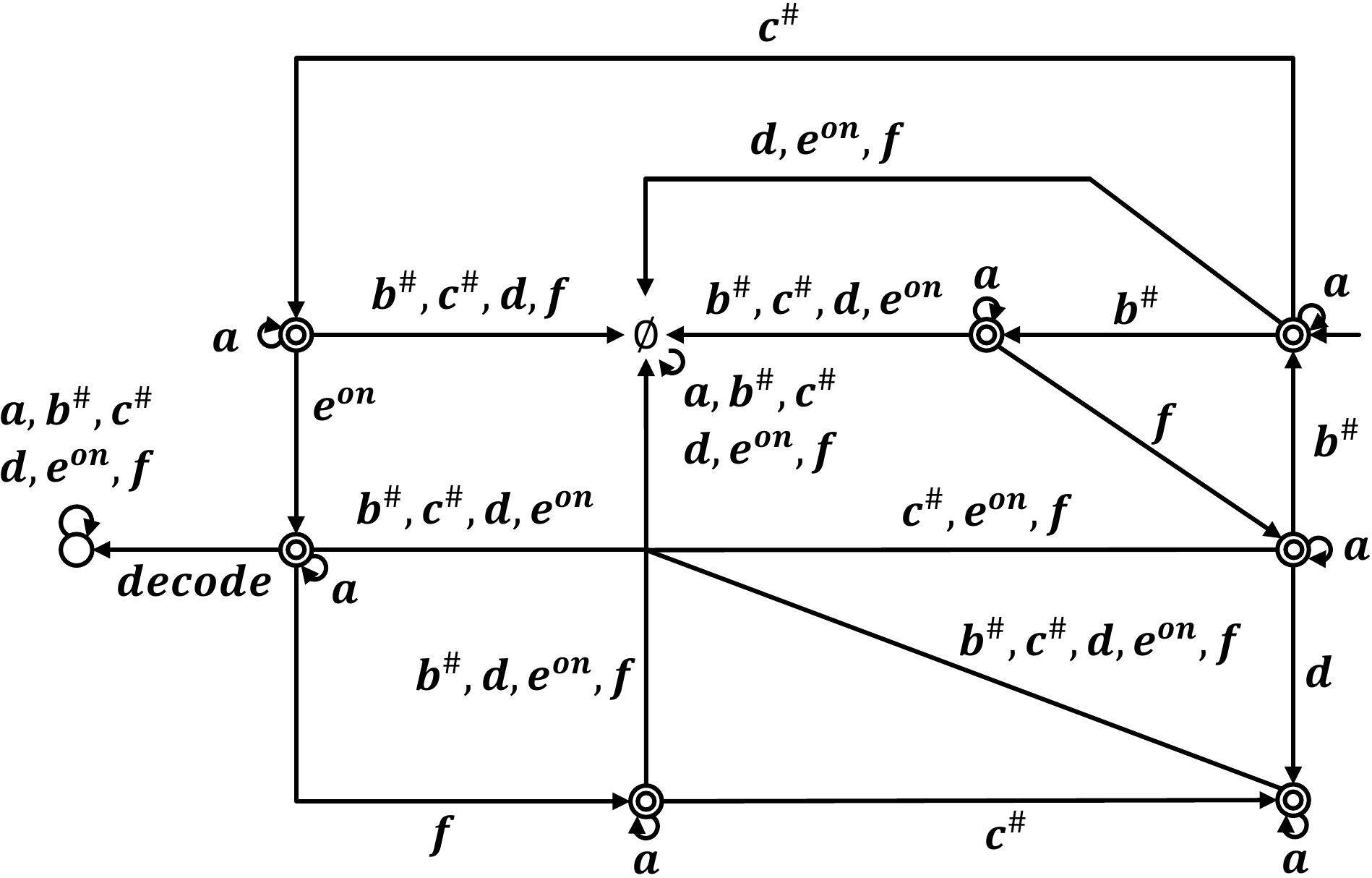}   
\caption{Sensor eavesdropping intruder $I^{w}$}
\label{fig:Weak intruder I^w}
\end{center}        
\end{figure}

Plant $G$ and its requirement are shown in Fig. \ref{fig:Plant and Requirement}, where the state 5 is the secret state. 
Command execution automaton $CE$ is shown in Fig. \ref{fig:Command execution CE}. For mask constraints $MC = MC_{3}||MC_{5}$, $MC_{3}$ and $MC_{5}$ are shown in Fig. \ref{fig:Mask constraints MC}. Edit constraints $EC$ is shown in Fig. \ref{fig:Edit constraints EC}. Supervisor constraints $SC$ is shown in Fig. \ref{fig:Supervisor constraints SC}. The sensor-command eavesdropping intruder $I$ is shown in Fig. \ref{fig:intruder I}, where all the events in $\{a,b^{\#},c^{\#},d,e^{on},f, v_{1},v_{2},v_{3},v_{4},v_{5},v_{6},v_{7},\}$ should be self-looped at the state $\varnothing$, and for the sake of 
making the figure look clear, we only draw the self-loops of those events at the state $\varnothing$ in the upper left corner of Fig. \ref{fig:intruder I}. The sensor eavesdropping intruder $I^{w}$ is shown in Fig. \ref{fig:Weak intruder I^w}. 
%\subfigure[]{

%\begin{figure}[htbp]
%\begin{center}
%\includegraphics[height=4.3cm]{fig/ME.pdf}   
%\caption{Synthesized $ME$}
%\label{fig:Synthesized ME}
%\end{center}        
%\end{figure}

\begin{figure}[htbp]
\begin{center}
\includegraphics[height=9.8cm]{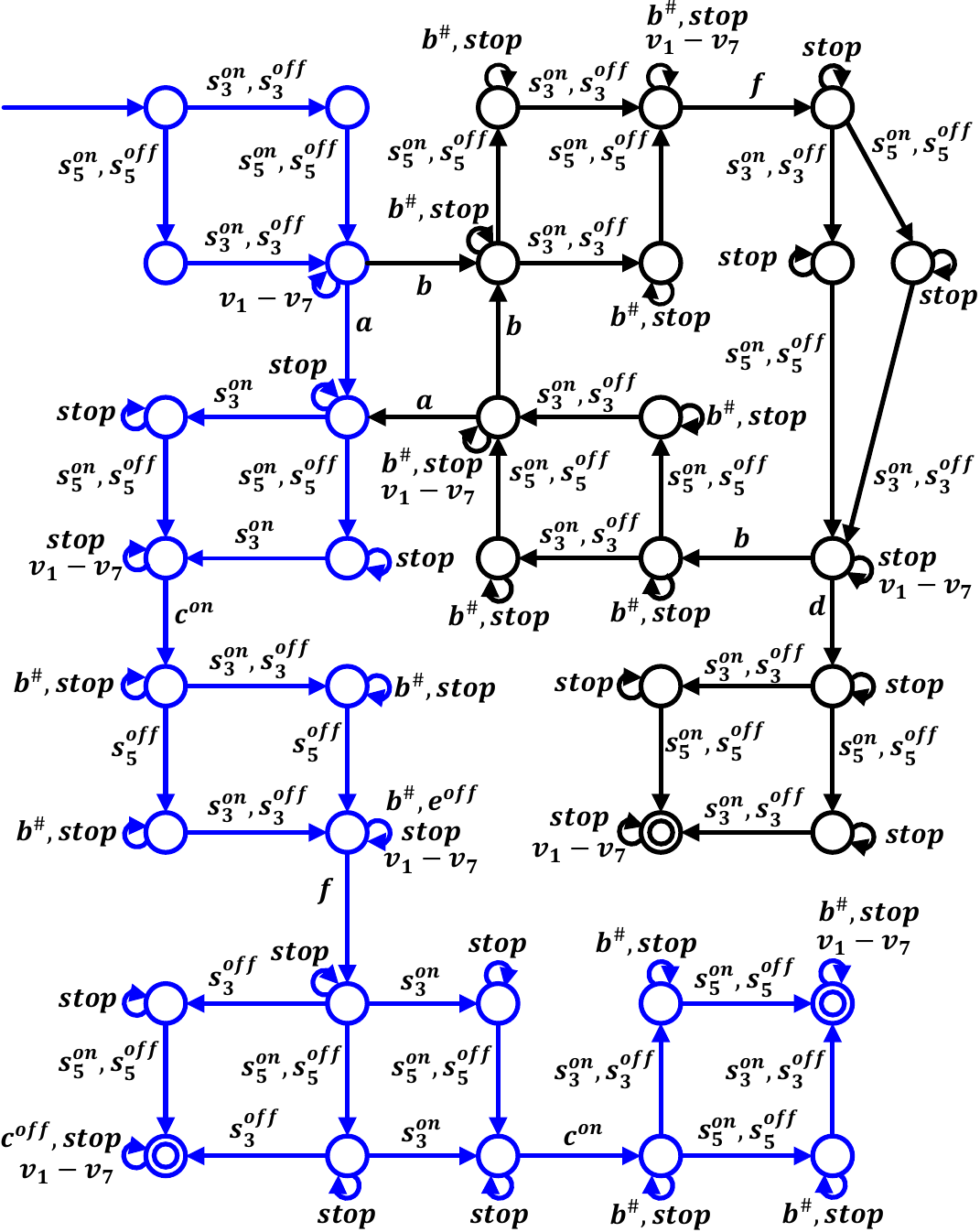}   
\caption{Synthesized $M$}
\label{fig:Synthesized M}
\end{center}        
\end{figure}

\begin{figure}[htbp]
\begin{center}
\includegraphics[height=7.3cm]{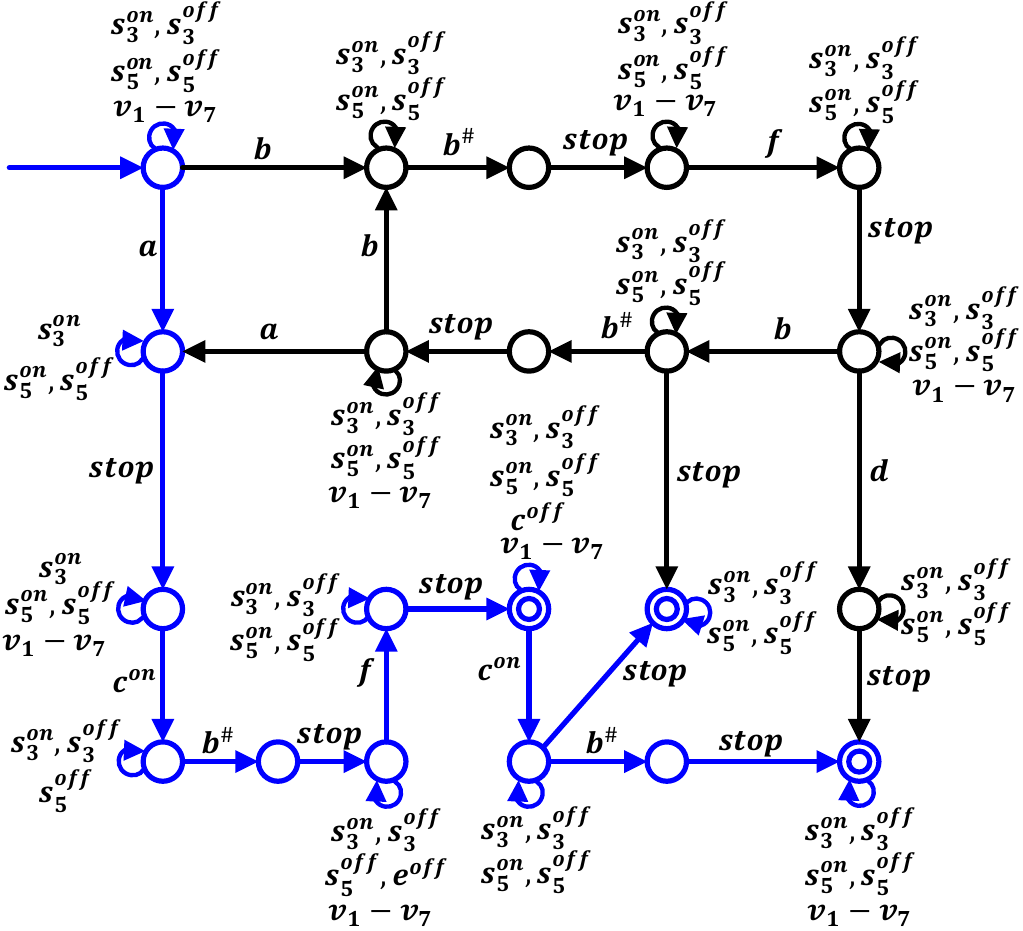}   
\caption{Synthesized $E$}
\label{fig:Synthesized E}
\end{center}        
\end{figure}

\begin{figure}[htbp]
\begin{center}
\includegraphics[height=3.5cm]{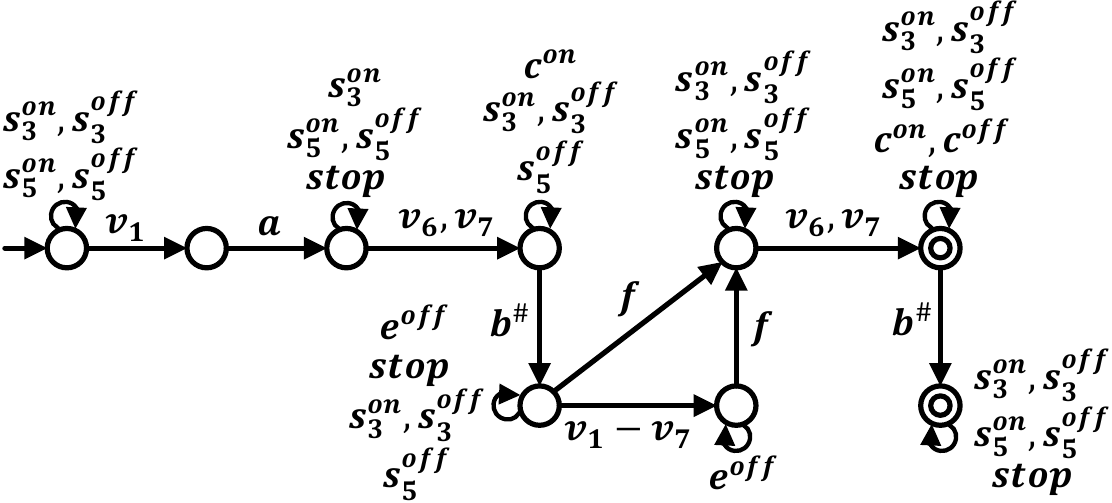}   
\caption{Synthesized $S$}
\label{fig:Synthesized S}
\end{center}        
\end{figure}

Based on \textbf{Procedure 1}, we adopt SuSyNA \cite{SuSyNA} to compute $ME$ and $S$. The decomposition of $ME$ into an $M$ and an $E$ can be computed by a SAT solver.
The synthesized $M$, $E$, and $S$ are shown in Fig. \ref{fig:Synthesized M}, \ref{fig:Synthesized E}, and \ref{fig:Synthesized S}, respectively, where the blue highlighted parts in Figs.\ref{fig:Synthesized M} and \ref{fig:Synthesized E} are the effective mask operations and edit operations under the cooperation of these three components. 
%{\color{red} I here do not understand}. 
It can be checked that the synthesized result can achieve the following goals: 1) the closed-loop system behavior is nonblocking and satisfy the requirement; 2) the intruder could never infer that plant $G$ has reached the secret state; 3) the dynamic mask and the edit function always remain covert. Next, we shall present some explanations for this result.

The intuitive strategy deployed by the synthesized result is that the plant $G$  executes the string $acefc$ and then the dynamic mask and the edit function will change such information flow to $bfb$ to deceive the intruder such that opacity is ensured, meanwhile the supervisor will cooperate with the dynamic mask and the edit function to issue appropriate control commands, which could be eavesdropped by the intruder, such that the covertness could be maintained. The details of the strategy are as follows:
At the initial state, $M$ could either turn on or turn off the sensor $s_{3}$ and $s_{5}$. Then $S$ issues the initial control command $v_{1}$, which cannot be eavesdropped by the intruder. After receiving $v_{1}$, $G$ executes event $a$, which is unobservable to the intruder. When $M$ observes $a$, it would turn on the sensor $s_{3}$ and either turn on or turn off the sensor $s_{5}$.
Since $a \notin \Sigma_{s,E}$, $E$ cannot change $a$. Then, the event $a$ will be observed by $S$, triggering it to issue the control command $v_{6}$ or $v_{7}$. When $G$ receives $v_{6}$ or $v_{7}$, it executes event $c^{on}$. After observing $c^{on}$, $M$ would turn off the sensor $s_{5}$ and either turn on or turn off the sensor $s_{3}$. The observation of $c^{on}$ triggers $E$ to change it to $b^{\#}$. Since $v_{6}$ or $v_{7}$ could be observed by the intruder and anyone of them contains the event $b$ and $c$, the sensor information alteration from $c^{on}$ to $b^{\#}$ would not expose the existence of $M$ and $E$. After that, $b^{\#}$ would be observed by $S$. Then, depending on whether the uncontrollable events $e$ and $f$ are executed by plant $G$ before $S$ issues a new control command, there are three cases:
\begin{enumerate}[1.]
\setlength{\itemsep}{3pt}
\setlength{\parsep}{0pt}
\setlength{\parskip}{0pt}
    \item $G$ sequentially executes $e^{off}$ and $f$ without receiving any control command, which is before $S$ issues a new control command.  
    \item $G$ executes $e^{off}$ without receiving any control command. After that, $S$ issues any of $v_{1}$-$v_{7}$, which would be received by $CE$, and event $f$ is executed by the plant $G$.
    \item $S$ issues any of $v_{1}$-$v_{7}$, which will be received by $CE$, resulting in the execution of the event $e^{off}$ at the plant $G$. After that, $G$ executes $f$.  
\end{enumerate}
When $M$ observes $f$, it would turn off (on, respectively) the sensor $s_{3}$ and either turn on or turn off the sensor $s_{5}$, then the observation of $f$ triggers $S$ to issue the control command $v_{6}$ or $v_{7}$, resulting in the execution of $c^{off}$ ($c^{on}$, respectively) at the plant $G$. If the sensor $s_{3}$ is turned on, then after observing $c^{on}$, $E$ could either delete $c^{on}$ or change $c^{on}$ to $b^{\#}$. Since $v_{6}$ or $v_{7}$ can be eavesdropped by the intruder and events $b$ and $c$ are both contained in $v_{6}$ or $v_{7}$, such a sensor information alteration would still ensure the covertness. Thus, what the intruder observes  could be $v_{6}/v_{7}b^{\#}fv_{6}/v_{7}$, or $v_{6}/v_{7}b^{\#}fv_{6}/v_{7}b^{\#}$, or $v_{6}/v_{7}b^{\#}v_{i}(i \in [2,7])fv_{6}/v_{7}$, or $v_{6}/v_{7}b^{\#}v_{i}(i \in [2,7])fv_{6}/v_{7}b^{\#}$, which could ensure that the opacity and covertness are both satisfied. Moreover, it can be checked that both of the dynamic mask and the edit function are indispensable to ensuring the opacity and the covertness in this example, which  illustrates the necessity of adopting the dynamic mask and the edit function together for opacity enforcement.

%%%%%%%%%%%%%%%%%%%%%%%%%%%%%%%%%%%%%%%%%%%%%%%%%%%%%%%%%%%%%%%%%%%%%%%%%%%%%%%%

\section{Conclusions}
\label{sec:conclusions}
In this paper, we address the privacy-preserving supervisory control problem for opacity enforcement and requirement satisfaction, where the intruder could eavesdrop both the sensing information and control commands. By modeling this problem as a distributed supervisor synthesis problem in the Ramadge-Wonham supervisory control framework, our methods allow existing synthesis tools such as SuSyNA \cite{SuSyNA}, Supremica \cite{akesson2006supremica}, or TCT \cite{feng2006tct} to be used. 
In our future works, we shall explore other distributed synthesis approaches to generate potentially better distributed solutions. 

%\begin{ack}                               % Place acknowledgements
%The funding supports from Singapore National Research Foundation via Delta-NTU Corporate Lab Program (DELTA-NTU CORP
%LAB-SMA-RP2 SU RONG M4061925.043) and from Singapore Ministry of Education Tier
%1 Academic Research Grant (M4011982 RG91/18-(S)-SU RONG (VP)) are gratefully
%acknowledged.  % here.
%\end{ack}

%\vspace{0.4cm}

\begin{IEEEbiography}[
{
\includegraphics[width=1.0in,height=1.40in,clip,keepaspectratio]{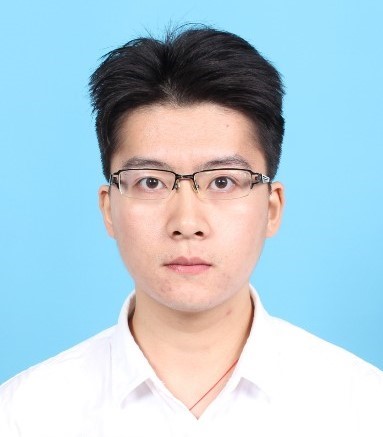}
}
]
{Ruochen Tai}
received the B.E. degree in electrical engineering from the Nanjing University of Science and Technology in 2016, and M.S. degree in automaton from the Shanghai Jiao Tong University in 2019. He is currently pursuing the Ph.D. degree with Nanyang Technological University, Singapore. His current research interests include security issue of cyber-physical systems, multi-robot systems, safe autonomy in cyber-physical-human systems, formal methods, and discrete-event systems.
%systems.
\end{IEEEbiography}
\begin{IEEEbiography}[
{
\includegraphics[width=1.0in,height=1.40in,clip,keepaspectratio]{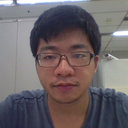}
}
]
{Liyong Lin}
received the B.E. degree and Ph.D. degree in electrical engineering in 2011 and 2016, respectively, both from Nanyang Technological University, where he has also worked as a project officer. From June 2016 to October 2017, he was a postdoctoral fellow at the University of Toronto. Since December 2017, he has been working as a research fellow at the Nanyang Technological University. His main research interests include supervisory control theory, formal methods and machine learning. He previously was an intern in the Data Storage Institute, Singapore, where he worked on single and dual-stage servomechanism of hard disk drives.
%systems.
\end{IEEEbiography}
\begin{IEEEbiography}[
{
\includegraphics[width=1.0in,height=1.20in,clip,keepaspectratio]{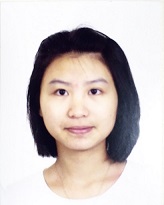}
}
]
{Yuting Zhu}
received the B.S. degree from Southeast University, Jiangsu, China, in 2016. She is currently pursuing the Ph.D. degree with Nanyang Technological University, Singapore. Her research interests include networked control and cyber security of discrete event systems.
%systems.
\end{IEEEbiography}
\begin{IEEEbiography}[
{
\includegraphics[width=1in,height=1.3in,clip,keepaspectratio]{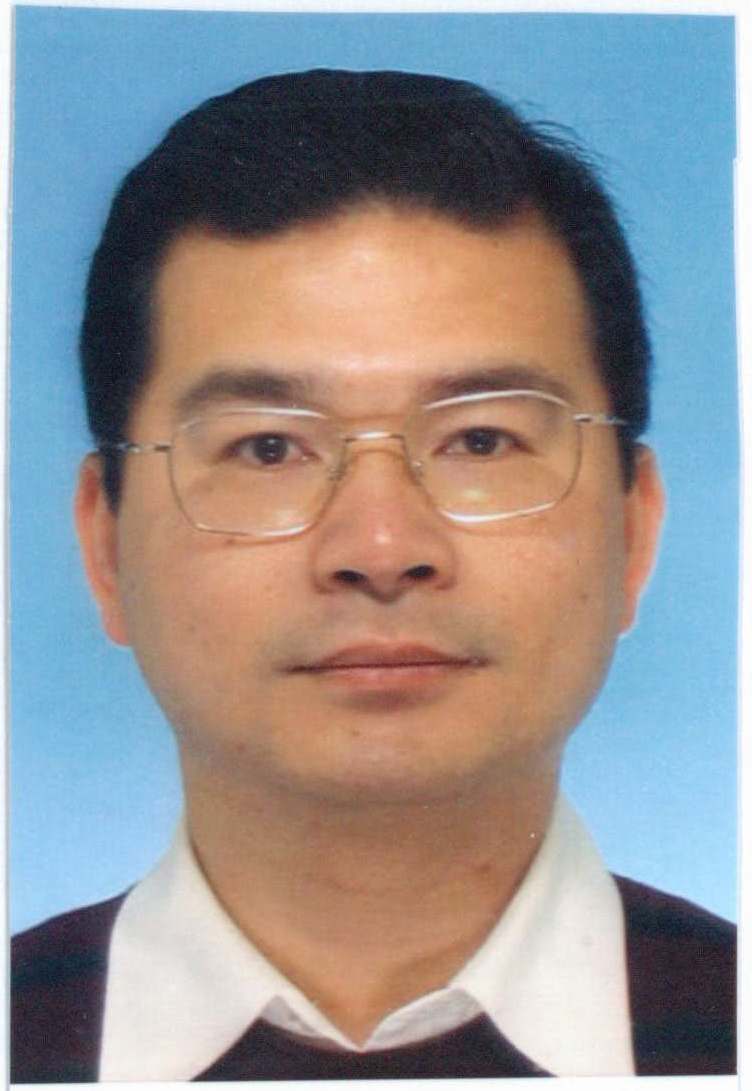}
}
]
{Rong Su} received the Bachelor of Engineering degree from University of Science and Technology of China in 1997, and the Master of Applied Science degree and PhD degree from University of Toronto, in 2000 and 2004, respectively. He was affiliated with University of Waterloo and Technical University of Eindhoven before he joined Nanyang Technological University in 2010. Currently, he is an associate professor in the School of Electrical and Electronic Engineering. Dr. Su's research interests include multi-agent systems, cyber security of discrete-event systems, supervisory control, model-based fault diagnosis, control and optimization in complex networked systems with applications in flexible manufacturing, intelligent transportation, human-robot interface, power management and green buildings. In the aforementioned areas he has more than 220 journal and conference publications, and 5 granted USA/Singapore patents. Dr. Su is a senior member of IEEE, and an associate editor for Automatica, Journal of Discrete Event Dynamic Systems: Theory and Applications, and Journal of Control and Decision. He was the chair of the Technical Committee on Smart Cities in the IEEE Control Systems Society in 2016-2019, and is currently the chair of IEEE Control Systems Chapter, Singapore, and a co-chair of IEEE Robotic and Automation Society Technical Committee on Automation in Logistics.

\end{IEEEbiography}

\end{document}